\documentclass[numbers]{cas-sc}
\usepackage{graphicx}
\usepackage{amsmath,amsthm, amssymb, float, bm, subcaption}
\usepackage{placeins}
\usepackage{longtable,tabularx,booktabs}
\usepackage{soul}
\usepackage{tcolorbox}
\tcbuselibrary{skins}
\usepackage{xcolor}
\usepackage[percent]{overpic}
\usepackage{miller}

\definecolor{C0}{HTML}{1F77B4}
\definecolor{C1}{HTML}{FF7F0E}
\definecolor{C2}{HTML}{2ca02c}
\definecolor{C3}{HTML}{d62728}
\definecolor{C4}{HTML}{9467bd}
\definecolor{C5}{HTML}{8c564b}

\newtcolorbox{mybox}[1][]{colback=#1!10,colframe=#1!80,boxrule=1pt,arc=1mm,left=0.1mm,right=0.1mm,top=0.1mm,bottom=0.1mm}

\usepackage[capitalize]{cleveref}

\makeatletter
\define@key{Gin}{Crop3D}[]{\setkeys{Gin}{trim=21mm 15mm 80mm 22mm,clip}}
\makeatother

\makeatletter                                                                   
\newlength{\bibsep}{\@listi \global\bibsep\itemsep \global\advance\bibsep by\parsep} 
\makeatother 

\makeatletter
\def\ps@pprintTitle{%
  \let\@oddhead\@empty
  \let\@evenhead\@empty
  \def\@oddfoot{\reset@font\hfil\thepage\hfil}
  \let\@evenfoot\@oddfoot
}
\makeatother

\begin{document}

\ifdefined\usetodonotes
\setcounter{page}{0}
\listoftodos
\fi

\shorttitle{Predicting void nucleation in microstructure with convolutional neural networks}
\shortauthors{Anantharanga, Plummer, Fensin, Runnels}  
\title [mode = title]{Predicting void nucleation in microstructure with convolutional neural networks}  
\author[isu]{Abhijith Thoopul Anantharanga}
\author[isu]{Jackson Plummer}
\author[lanl]{Saryu Fensin}
\author[isu]{Brandon Runnels}[orcid=0000-0003-3043-5227]
\cormark[1]
\cortext[1]{Corresponding author}
\ead{brunnels@iastate.edu}
\affiliation[isu]{organization={Department of Aerospace Engineering, Iowa State University},
city={Ames},
state={IA},
country={USA}}
\affiliation[lanl]{organization={Los Alamos National Laboratory},
            city={Los Alamos},
            state={NM},
            country={USA}}

\begin{abstract}
Void nucleation in ductile materials subjected to high strain-rate loading remains a critical yet elusive phenomenon to understand.   
Traditional methods to understand void nucleation typically rely on experiments and molecular dynamics and do not capture the underlying factors leading to void nucleation.  
In this study, a convolutional neural network, specifically a U-Net enhanced with attention gates is developed, to predict void nucleation probability in pristine tantalum microstructures.
The approach leverages a multi-channel input, incorporating four channels of grain orientations and an additional channel of grain boundary energy calculated via the lattice matching method. 
Void nucleation probability fields are determined from {\it post-mortem} micrographs and serve as ground truth, distinguishing void from no-void regions at the pixel level. 
Pixel-level class imbalance, commen in such images, is addressed by using Focal loss to guide the network’s training to predict void nucleation sites more effectively.
The model not only predicts void nucleation sites consistent with ground-truth but also reveals additional potential void nucleation sites, capturing the stochastic nature of void nucleation.  
This study shows that CNN-based models can predict void nucleation sites while considering combined interplay of factors such as grain boundary energy and grain orientation.
In this way, machine learning can serve as a means to understand the underlying factors leading to void nucleation thereby contributing to a fundamental understanding of failure due to spallation in ductile materials.
\end{abstract}

\maketitle

\section{Introduction}

Ballistic impacts, collisions of space debris with spacecraft, and other high-velocity events produce dynamic loads which subject materials to extreme strain rates \cite{remington2015microjoules}.
Ductile materials subjected to high strain rates primarily fail via spallation  \cite{davison1977theory}.
Spallation is a series of complex phenomena that include void nucleation, growth, coalescence, and ultimately separation of material that lead to failure during high impact dynamic loads under a high strain rate \cite{chevrier1999spall}.
Under shock loading, such as from plate impact or detonation, a material is under the influence of a compressive shock. 
When the compressive shock waves reach a free surface of the material, they reflect back as rarefaction waves.
The interaction of opposing rarefaction fans within the sample can generate a localized pulse of extreme tension\cite{antoun2003spall}.
If the magnitude of this tensile pulse is greater than the critical stress the voids can nucleate in ductile materials. 
When these voids nucleate, grow, and coalesce, spall failure occurs.
Incipient spall is early stage spallation in which voids nucleate and grow but do not coalesce and cause failure.
While all materials ultimately fail under sufficiently extreme conditions, designing microstructures to resist void nucleation and delay catastrophic failure due to spallation can significantly enhance performance.

Experiments on incipient spall have found that voids tend to nucleate predominantly at grain boundaries in ductile materials. \cite{fensin2014some}
It has been identified experimentally that boundaries with a high $\Sigma$ orientation relationship tend to preferentially nucleate voids. \cite{escobedo2011effects}
Experimental observations of face-centered cubic (FCC) copper find that some boundaries, namely the $\Sigma 3$ and low-angle boundaries, tend to resist void nucleation \cite{cerreta2012early}.
Other experiments find that the tips of terminated twin boundaries are common sites for void nucleation \cite{brown2015microstructural}.
Molecular dynamics (MD) simulations on copper have indicated that voids are more likely to nucleate at grain boundaries oriented perpendicular to the impact-stress direction \cite{yang2022grain}. 
Further, MD simulations to understand the relationship between spall strength and grain boundary properties infer that there is no relationship between grain boundary energy and excess free volume with spall strength in $\Sigma3$, $\Sigma5$, $\Sigma9$, $\Sigma11$ ordered and $\Sigma11$ disordered boundaries. \cite{fensin2012influence}.

Studies on void nucleation also seek to make connections between factors other than grain boundaries.
Misorientation angles between 25$^{\circ}$ and 50$^{\circ}$ have been found to nucleate voids \cite{wayne2010statistics}.
Molecular dynamics (MD) simulations corroborate these findings in tantalum \cite{Homer:2022:AlGBdataset}.
Voids have been observed to form along grain boundaries oriented parallel, perpendicular, or 45 degrees to the tensile axis, indicating a relationship between void formation and grain boundary orientation relative to the applied stress \cite{lillo2009influence}.
However, the role of misorientation angle as a grain boundary descriptor has been called into question as experimental evidence suggests it may be an insufficient predictor for void nucleation on its own\cite{nguyen2019role}.
Experimental work to understand the relationship between microstructure and spall fracture in aluminum shows that crystallographic orientation, relative to the direction of shock propagation, does not discernibly affect spall strength in Al-HP material and the grain size in  Al–3Mg material does not affect spall strength \cite{pedrazas2012effects}.
Further studies on the effect of grain size on void nucleation due to shock loading indicate that grain size has no visible effects in FCC copper whereas body-centered cubic (BCC) tantalum not only shows voids along intergranular boundaries but also has voids developing inside of voids as grain size increases \cite{chen2019understanding}.
Collectively, these studies show that while certain microstructural features like grain boundary character, orientation, and misorientation, can promote void nucleation; none of them alone are reliable predictors. 
Many boundaries that fail share these necessary characteristics, yet others with the same attributes remain intact.
This raises a fundamental question: why do some boundaries fail, but some others do not?

Despite significant attempts via experiments and MD simulations, this question has not been answered and identifying the underlying mechanisms for void nucleation remains a challenging task. 
In this work, it has been hypothesized that void nucleation is driven by a complex interaction of grain boundary properties, crystallographic orientation, and the local environment. 
Given this complex interplay, the role of machine learning (ML) becomes paramount in the framework used in this work to understand the interplay between these various factors.
The adoption of ML enables us to overcome the limitations of experiments and MD simulations, offeing a more nuanced and accurate approach to detecting potential void nucleation sites in microstructures.
Using ML, it is possible to automate the process of predicting void nucleating sites which is impossible for the human eye to predict by visually inspecting a microstructure. 
Recent advances in ML provide ways to capture patterns within high-dimensional datasets and offer a data-driven framework to understand the subtle correlations between input data like grain orientations and grain boundary energies that contribute to void nucleation. 

The application of ML to materials science has shown demonstrated success in tasks such as fracture detection, fatigue prediction, and constitutive modeling \cite{bock2019review, liu2020machine}.
Among ML architectures, multilayer perceptrons (MLPs) \cite{taud2018multilayer} and variational autoencoders (VAEs) \cite{anantharanga2023linking} have been utilized to map complex relationships in material design problems, while convolutional neural networks (CNNs) have enabled automated defect classification and microstructural image segmentation \cite{o2015introduction, pinz2022data, long2015fully}. 
The U-Net architecture, originally developed for biomedical image segmentation \cite{ronneberger2015u}, has emerged as a powerful tool for pixel-wise segmentation tasks.
Further, attention mechanisms are incorporated in U-Net architectures to enhance the predictive capabilities of the model \cite{oktay2018attention}
In conventional U-Net applications, the network learns to segment features by training on input images paired with pixel-level annotated masks. However, this study introduces a novel adaptation of the U-Net architecture with attention mechanism: instead of using direct pixel-wise mask as training labels, the model is trained solely on input features with pixel-wise fields being introduced in the loss function computation. 
The aim is to learn a mapping that predicts the probability of void nucleation across the microstructure using inputs to the U-net, without ever directly exposing the model to post-spall void locations via input data.

The goal of this work is to gain foundational insights into void nucleation and close existing knowledge gaps in understanding void nucleation in microstructures under incipient spallation using ML.
In this work, 57 post-incipient spall micrographs obtained from plate impact experiments are analysed to predict void nucleation probability.
Towards this end, these {\it post-mortem} micrographs are reconstructed to obtain pristine microstructures.
The lattice matching method is then incorporated to calculate the grain boundary energies of all grain boundaries within these reconstructed microstructures. 
Grain orientation information from Electron Backscatter Diffraction (EBSD) data, combined with computed grain boundary energies and local microstructural environments are used as inputs to an ML model which then predicts void nucleation probability fields by considering cumulative effects of all inputs to the ML model. 
The structure of the paper is as follows.
The second section describes techniques for reconstructing microstructure from experimental data that are later processed using the ML model (\cref{sec:MCR}). 
The main focus of this work is explored in the section, describing the development of a neural network aimed at predicting void nucleation (\cref{sec:nn}). 
The results section showcases the model's effectiveness in accurately identifying void nucleation sites (\cref{sec:results}).
The paper concludes with the conclusions section, highlighting findings, reflecting on their implications for void nucleation, and suggesting future research avenues. (\cref{conclusions}).

\section{Dataset generation}\label{sec:MCR}

Plate impact experiments were conducted at Los Alamos National Laboratory on BCC tantalum specimen. 
A review of the experiment is presented, but the reader is referred to \cite{jones2018effect} for more details.
Tantalum used in this study was sourced from H.C. Starck GmbH as a 50 cm diameter, 1 cm thick plate. 
To ensure a uniform microstructure, it underwent clock-rolling, asymmetric tilt-rolling, and heat treatment. 
The samples were then sectioned such that the shock-loading direction aligned with the through-thickness direction, after which plate-impact tests were conducted using an 80 mm bore single-stage light gas gun.
The recovered samples were cut diametrically with a diamond saw, mounted in epoxy, polished through successive grinding steps, and lightly acid-etched to reveal microstructural features.  
Optical micrographs were imaged using a Zeiss AX10 microscope and processed using ImageJ.
Because optical micrographs can only be used to characterize visible voids and damage, EBSD was used to study sample cross-sections and to obtain information about the crystallographic orientation of grains and end-point coordinates of each grain.
This combination of experiments and post-processing by EBSD produced 60 {\it post-mortem} micrographs along with processed EBSD data associated with each micrograph. 

The use of the 2D experimental characterization results enables the generation of adequate data to train and test a robust machine learning model, beyond what would be achievable with full 3D characterization.
However, the restriction to 2D also introduces two potential source of uncertainty due to the loss of information incurred by considering 2D slices only.
First, 2D slices that contain voids are (in principle) unable to distinguish voids that are small, and sub- or super-surface voids that were sliced eccentrically.
While it is true that such error exists in 2D data, a simple calculation shows that the average measured radius $\langle A \rangle$ of a pore with radius $A$ intersected by a uniform distribution of planes is $\langle A \rangle \approx 0.8 A$.
That is, the {\it measured} pore radius is, on average $80\%$ of the {\it actual} pore radius.
This is accounted for in the microstructure reconstruction process.
The second restriction of 2D is the loss of the out-of-plane inclination information.
This is discussed below in the calculation of GB energy.

\subsection{Microstructure reconstruction}

Voids in {\it post-mortem} micrographs are easily visible to the naked eye.
However, the features leading to void nucleation exist in pristine, undamaged microstructure, and so any void prediction model must be based exclusively on the microstructure features that exist prior to damage.
{\it In situ} or before-and-after EBSD characterization is not always possible or feasible at the length scale of interest, and may be too expensive to generate the size of dataset required for a ML model.
Therefore, a microstructure reconstruction technique is necessary to furnish training and testing data without leaving distinguishing processing features \cite{bostanabad2018computational}.

The use of reconstructed microstructure limits the available data to regions in which no significant damage has taken place.
Large experimental datasets must be partitioned into subregions that contain only regions with either no visible damage, or minimal damage that consists only of small, localized voids.
This is automated using a separate machine learning model and a partitioning algorithm.
(Additional details are provided in \cref{sec:dataset_preparation} and supplementary material.)

\begin{figure}
    \centering
    \includegraphics[height=8cm]{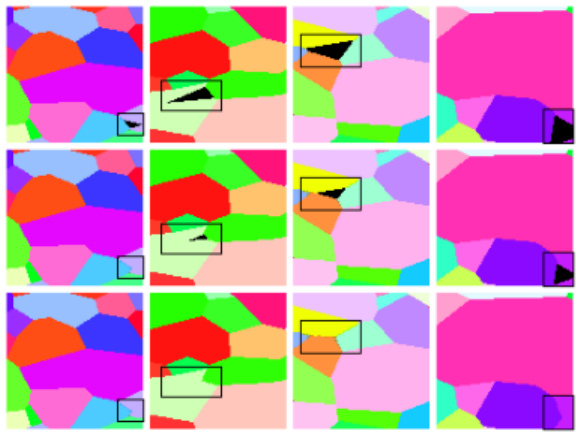}
    \caption{
    A reversed erosion algorithm, together with EBSD and grain boundary information, is used to reconstruct pre-damaged microstructures.
    (top) Samples of EBSD data showing defects resulting from the presence of voids.
    (middle) Erosion algorithm fills the defects until
    (bottom) The defect is completely filled.
    }
    \label{fig:erode}
\end{figure}

Following void detection and microstructure segmentation, GB information is used to back-fill EBSD regions, and a reverse-erosion algorithm is used on the EBSD maps to close void regions (\cref{fig:erode}).
After the boundary construction, grain orientations from EBSD derived Bunge Euler angles are visualized by grouping boundary endpoints for each grain, computing their centroid, and forming closed polygons.
Finally, a Gaussian blur is applied to the reconstructed grain orientation image to smooth out artifacts introduced during grain reconstruction and merging. 

\subsection{Grain boundary energy}

Grain boundary energy (GBE) is well-known to play a role in material failure.
Even though no simple correlations have yet been found between GBE and void nucleation, it is hypothesized that GBE is one of many factors leading to failure likelihood.
GBs are often called the ``least-understood'' of material defects due to their high dimensionality - the macroscopic configuration alone requires five degrees of freedom to characterize.
Numerous studies have been performed to calculate GBE for arbitrary GBs using atomistic methods \cite{hartford2000interface}, \cite{wright1994density},\cite{kang2012minimum} and interpolation \cite{bulatov2014grain}, but determination of GBE for arbitrary materials at arbitrary orientations remains challenging.
The lattice matching (LM) method provides a semi-analytical form for GBE, and has been shown to work for both FCC and BCC materials \cite{dobrovolski2023facet,jasperson2025fundamental}.
Its analytic form enables application to large datasets and phase field codes \cite{gokuli2021multiphase}, and the implementation has been openly released and well-tested \cite{wield}.
A brief overview of LM theory is presented here, but the reader is referred to \cite{runnels2016analytical} for an extended derivation.

The lattice matching approach is grounded in optimal transport theory, under the assumption that the grain boundary energy is proportional to the energetic cost of transforming one atomic lattice into another across an interfacial region. 
Consequently, the grain boundary energy is formulated as the minimizer of this transport cost, rendering the problem as one of optimal transport.
The central quantity in this framework is the \emph{lattice density field}, $\rho(\mathbf{x})$, which represents the distribution of atoms in a grain.
This enables the definition of the covariance functional along the boundary, defined as
\begin{equation}
c[\rho_1, \rho_2] = \frac{1}{\mu(1)} \int_{\partial \Omega} \rho_1(\mathbf{x}) \rho_2(\mathbf{x}) , d\mu(\mathbf{x}),
\label{eq:covariance_integral}
\end{equation}
where $\mu$ is the measure parameterized by a window function $\phi$.
By expanding the lattice density fields as Fourier series, the covariance admits a closed-form double sum expression:
\begin{equation}
c[\rho^1, \rho^2](\mathbf{R}^1, \mathbf{R}^2) = \frac{1}{\hat{\phi}(\mathbf{0})} \sum_{\mathbf{n}^1, \mathbf{n}^2 \in \mathbb{Z}^3} \rho{^1_{\mathbf{n}^1}} \rho{^2_{\mathbf{n}^2}} \hat{\psi}(\mathbf{k}_{\mathbf{n}^1}) \hat{\psi}(\mathbf{k}_{\mathbf{n}^2}) \hat{\phi}(\mathcal{P}(\mathbf{R}^1 \mathbf{k}_{\mathbf{n}^1} - \mathbf{R}^2 \mathbf{k}_{\mathbf{n}^2})),
\label{eq:covariance_sum}
\end{equation}
where $\mathbf{R}_1,\mathbf{R}_2\in SO(3)$,  $\hat{\phi}$ is the Fourier transform of $\phi$, $\hat{psi}$ is the Fourier transform of the atomic positions smeared out by temperature, and $\mathcal{P}$ is the projection operator onto the boundary plane.
The Fourier coefficients $\rho^{1}, \rho^{2}$ are lattice-specific and are precomputed for any lattice structure.
The GB covariance is then converted into a grain boundary energy $\gamma$ through the relationship:
\begin{equation}
\gamma(\mathbf{R}^1, \mathbf{R}^2) = \gamma_0 \left(1 - \frac{c(\mathbf{R}^1, \mathbf{R}^2)}{c_0} \right),
\label{eq:affine_gamma}
\end{equation}
where $c_0$ is the ground-state covariance, which is the covariance of the most densely packed plane, and $\gamma_0$ is the energy scaling factor which is typically determined by atomistic or {\it ab initio} data, and is constant for all GB orientations.
Evaluation of \cref{eq:affine_gamma} is sufficiently fast to feasibly calculate energies for all boundaries with minimal computational cost.

It is noted that, without out-of-plane orientation, GB energy calculation lacks the requisite five degrees of freedom to completely characterize the GB character.
This so-called \textit{fifth degree of freedom problem} has long been recognized as a major barrier in accurately characterizing GBs using 2D data. 
To address this, it is assumed that the grain boundaries are orthogonal to the plane. 


\subsection{Void nucleation probability maps}

\begin{figure}
    \centering
    \begin{minipage}[c]{0.15\linewidth}
      \vspace{0.5cm}
      \begin{mybox}[black]\textbf{GB energy}\end{mybox}
      \vspace{1.4cm} 
      \begin{mybox}[black]\textbf{Void mask}\end{mybox}
      \vspace{1.4cm} 
      \begin{mybox}[black]\textbf{Ground truth}\end{mybox}
      \vspace{0.5cm}
    \end{minipage}%
    \begin{minipage}[c]{0.85\linewidth}
      \includegraphics[width=\textwidth]{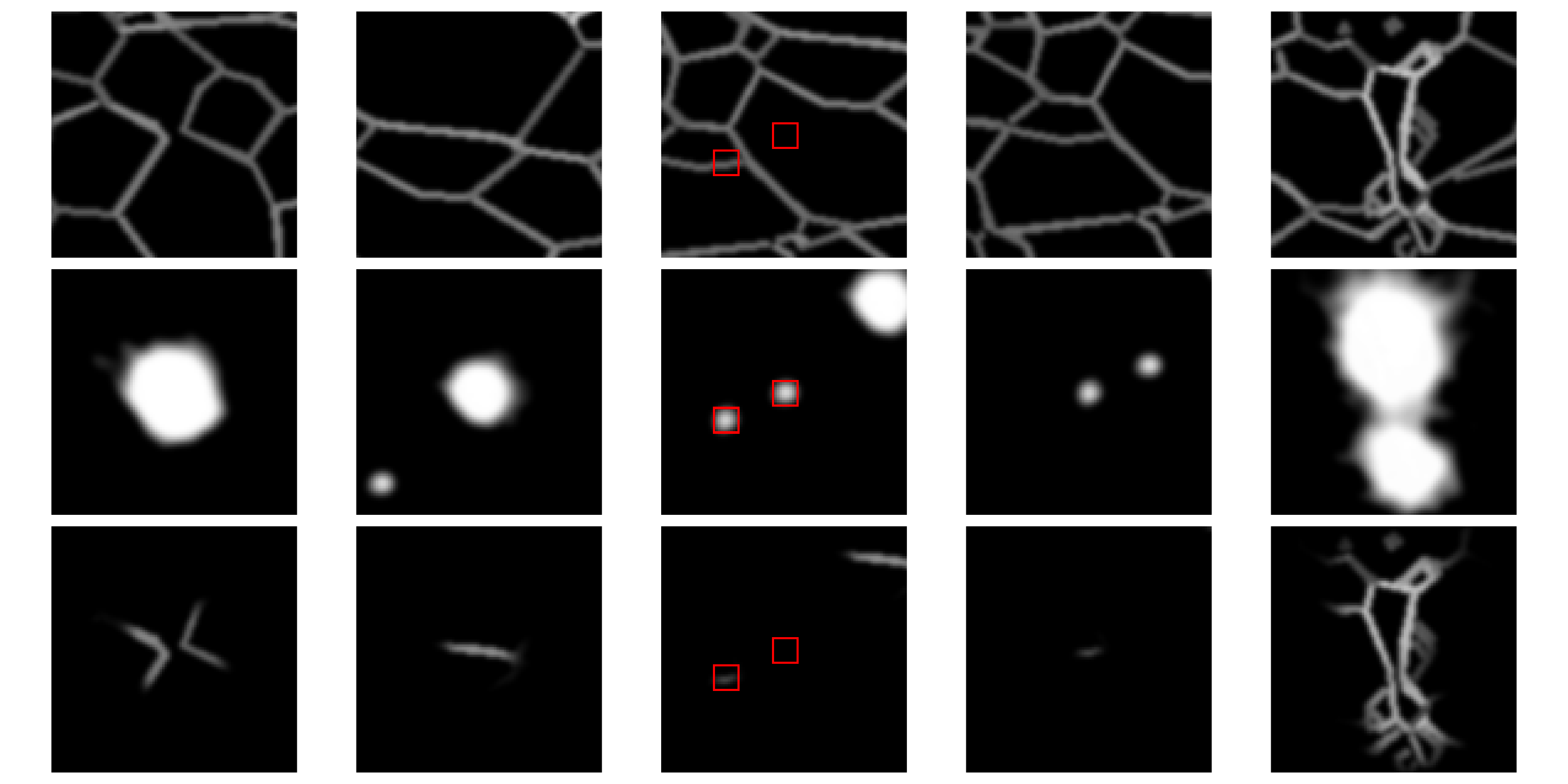}
    \end{minipage}
    \caption{A visual compilation of three image sets: grain boundary energy (top row), void masks (middle row), and final ground truth images (bottom row). The final ground truth images are obtained by performing a pixel wise multiplication of grain boundary energy and void masks. The red boxes emphasize two cases: (i) when voids intersect grain boundaries, resulting in retained regions in the final mask, and (ii) when voids fall outside grain boundaries, leading to their exclusion from the final mask. This approach ensures that only voids that are on grain boundaries are preserved for further analysis.}
    \label{fig:binary_mask}
\end{figure}

The desired outcome of the model is a map indicating the relative likelihood of void nucleation over the microstructure as a function of position.
For supervised learning, therefore, simulated maps must be created that approximate such likelihoods based on the observed void structure to serve as ``ground truth'' results.
A central hypothesis of this work is that voids nucleate at grain boundaries, which is consistent with the majority of observed voids.
This effect is approximated by the mask
\begin{align}
  D(x, y) = \tilde{M}(x, y) \cdot \tilde{G}(x, y)  = \Big(M*\mathcal{G}_\sigma\Big)\cdot\Big(G*\mathcal{G}_\epsilon\Big),
\end{align}
where $D$ represents the ground truth (void nucleation map), $\cdot$ is pointwise multiplication, $*$ is convolution, $\tilde{M}$ and $\tilde{G}$ are mollifications of $M$, $G$, which are the void map (from image segmentation) and the grain boundary energy (from lattice matching), respectively.
The mollifiers $\mathcal{G}$ are Gaussians with standard deviations $\sigma,\epsilon$ representing uncertainty in void location and grain boundary location, respectively.
The result reflects the possible locations of void nucleations {\it along pristine boundaries}, with uncertainty reflected through mollification (\cref{fig:binary_mask}).

It is emphasized that this method limits nucleation prediction to voids that may have nucleated along grain boundaries; voids that appear entirely in the interior of a grain are eliminated.
This accomplishes two things:
(1) it reduces the stochasticity of the dataset, and 
(2) it reduces the risk of spurious influence by voids that nucleated far from the cut surface.
This allows the model to ``focus'' on the specific effect of GBs, which is of primary interest in this work.
In practice, a separate nucleation criterion should be added to account for intragranular void nucleation.

\section{Machine learning model}\label{sec:nn}

After the dataset is assembled, an ML framework for predicting void nucleation in microstructures is developed.
As an initial step, data preparation is undertaken to ensure the dataset is thoroughly cleaned, normalized, and formatted for subsequent ML analyses. 
This step is critical for mitigating noise and inconsistencies, thereby enhancing model reliability. 
In the next phase, the choice of an ML model and its architecture is determined based on how suitable it is to predict void nucleation probability. 
The configuration of network layers, the selection of activation functions, and other hyper-parameters collectively impact the model’s capacity for learning complex features and generating accurate predictions.
Finally, the workflow culminates in an evaluation phase, wherein the model’s outputs are assessed to extract meaningful insights regarding void nucleation.
\begin{figure}
    \centering
    \begin{minipage}[c]{0.15\linewidth}
      \vspace{0.5cm}
      \begin{mybox}[black]\textbf{Micrograph}\end{mybox}
      \vspace{1.4cm} 
      \begin{mybox}[black]\textbf{GB energy}\end{mybox}
      \vspace{1.4cm} 
      \begin{mybox}[black]\textbf{Orientation}\end{mybox}
      \vspace{0.5cm}
    \end{minipage}%
    \begin{minipage}[c]{0.85\linewidth}
      \includegraphics[width=\textwidth]{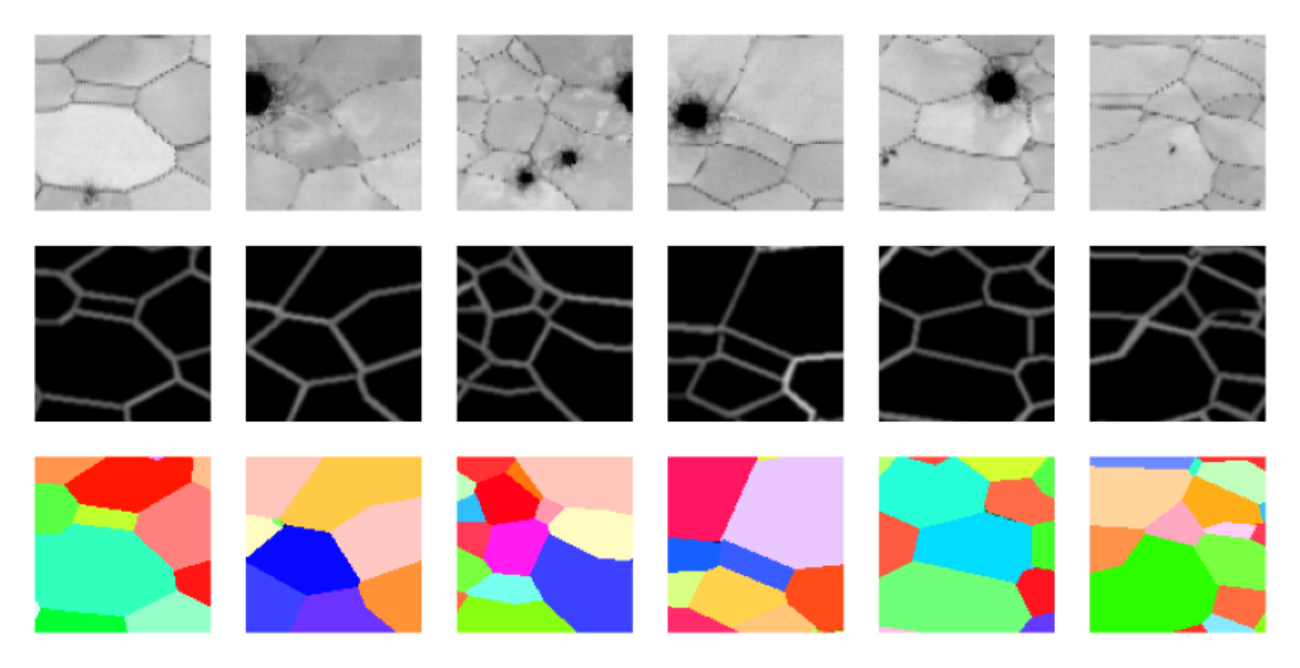}
    \end{minipage}
    \caption{A visual compilation of three image sets: micrograph (top row), grain boundary energy (middle row), and grain orientation (bottom row). The dataset to the ML model includes grain orientations and grain boundary energies.}
    \label{fig:grain_gb}
\end{figure}

The dataset comprises a set of 10326 images of microstructural images carrying the multichannel inputs consisting of grain orientation information and grain boundary energy (\cref{fig:grain_gb}), along with ground truth void nucleation fields. 
To ensure the prevention of gimbal lock, all angular representations of grain orientations from processed EBSD data are converted to quaternions from Euler angles. 
As a result, the grain orientation images are encoded with RGBA channels, each image with dimensions of 100x100x4.
Additionally, to simplify the representation of grain boundary energy, a single channel of 100x100x1 is used. 
The grain orientation images and the grain boundary energy images are concatenated to obtain 100x100x5 dimension data which is further processed via normalization to scale all values to a range between 0 and 1.
To ensure a comprehensive evaluation of the model, the leave-one-out cross validation technique is used. 
In each partition, all data originating from a single micrograph are assigned to the test set, while the data from the remaining 56 micrographs constitute the training set.
Repeating this procedure for every micrograph yields 60 independent train–test splits, enabling a rigorous evaluation of model behaviour across the full range of microstructural variability.

\begin{figure}
  \centering
  \includegraphics[width=0.95\linewidth]{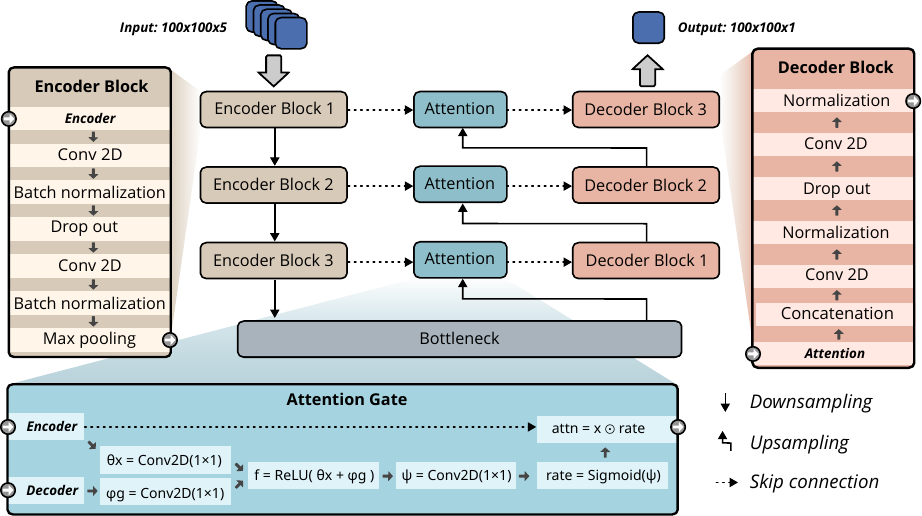}
  \caption{
  The full U-Net architecture, comprised of an encoder–decoder structure with attention gates between corresponding layers. The encoder processes a 100×100×5 input and the decoder reconstructs a 100×100×1 predictions.
  Popouts illustrate the encoder, decoder, and attention gate architectures.
  }
  \label{fig:mlmodel}
\end{figure}

\subsection{U-net machine architecture}

U-Net architectures have gained popularity for their success in segmentation tasks such as brain tumor detection \cite{mallampati2023brain}, farmland segmentation \cite{wang2023mde}, electron microscopy image segmentation \cite{cao2020denseunet}.
These largely disparate applications share the common goal of image segmentation and feature identification, which categorizes the present objective as well.
Even though regular autoencoders can be applied to similar tasks, they are often less accurate because they lack skip connections.
In a regular U-Net, information flows from the encoder to the decoder via the bottleneck layer. 
This indirect transfer pathway restricts the preservation of certain spatial features, leading to information loss. 
In contrast, U-Nets introduce skip connections that allow direct transfer of intermediate feature maps from the encoder to the decoder. 
These connections help retain feature information, significantly improving reconstruction quality and downstream task performance.

The model comprises a modified U-Net architecture tailored for the void nucleation probability field reconstruction from the processed and reconstructed microstructural input data (\cref{fig:mlmodel}). 
The network follows an encoder-decoder topology with symmetric skip connections between corresponding layers.
The encoder comprises three hierarchical blocks designed to progressively downsample the spatial resolution while expanding the number of feature channels.
Each block contains two 2D convolutional layers with a kernel size $3 \times 3$, stride 1, and same padding, followed by batch normalization and ReLU activations.
To improve regularization, a dropout layer (rate = 0.05) is applied after the first convolution in each block, and L2 weight decay (1e-4) is imposed on all convolutional kernels.
Downsampling is achieved by a $2 \times 2$ max-pooling operation at the end of each block.
The number of filters increases with depth, from 32 in the first block to 64 and 128 in the second and third blocks, respectively.
At the lowest resolution, the bottleneck captures high-level abstract features using two convolutional layers ( kernel size $3 \times 3$, stride 1, and same padding) with 256 filters, each followed by batch normalization and ReLU activation. 
A dropout layer is inserted after the first convolution to enhance robustness.
The decoder progressively restores spatial resolution while integrating contextual information from the encoder.
Each decoder block gets upsampled  by a factor of two, followed by a $2 \times 2$ convolution to refine features and reduce the number of channels.
Skip connections from the encoder are filtered through attention gates, ensuring that only the most salient encoder features contribute to reconstruction.
The gated encoder features are concatenated with the upsampled decoder features, and the combined maps are processed by convolutional blocks structurally identical to those in the encoder (two $3 \times 3$ convolutions with batch normalization, ReLU activations, and dropout after the first convolution).
The final output is generated using a $1 \times 1$ convolution followed by a sigmoid activation, producing a single-channel probability field prediction. 
The model improves the predicted void nucleation field by comparing them with the original ground truth void nucleation probability fields which are used in the loss function. 

The basic encoder/decoder U-net structure was shown to work well.
However, the inherent sparsity of voids in the data causes imbalance, which negatively affects the model performance. 
One way of addressing this is to deliberately prioritize void nucleation regions. 
This is accomplished through the incorporation of attention gates in the U-net, which adaptively highlights relevant regions of the encoder feature maps before they are fused with decoder features via skip connections.
Attention mechanisms have been previously used to mitigate issues which stem from severe class imbalance in applications such as road crack detection \cite{fan2022rao} and skin lesion classification \cite{wang2021deep}.
Various types of attention gates exist, such as hard attentions \cite{mnih2014recurrent}, recursive hard attentions \cite{ypsilantis2017learning}, soft attention \cite{bahdanau2014neural} and self attention \cite{wang2018non}.
In this work, the additive self-attention architecture similar to that used in \cite{oktay2018attention} was found to be the most effective. 
This is because this model incorporates additive attention \cite{bahdanau2014neural} to obtain the gating coefficients.
Although it is computationally more expensive, it has been shown to perform better than multiplicative attention \cite{luong2015effective}.
The structure of the gate is as follows: given an encoder feature map $x$ and a decoder gating signal $g$, both inputs are projected to an intermediate dimensionality through $1\times1$ convolutions, producing $\theta_x$ and $\phi_g$. 
These projected features are summed and passed through a ReLU activation to yield an intermediate feature map $f$, which is then reduced to a single-channel attention coefficient map $\psi$ via another $1\times1$ convolution.
A sigmoid activation is applied to $\psi$ to generate a spatial attention map $\text{rate} \in (0,1)$, which modulates $x$ through element-wise multiplication, producing the gated output $\text{attn} = x \odot \text{rate}$. 
This process effectively suppresses irrelevant background information and enhances salient structures, thereby improving segmentation accuracy while maintaining low computational overhead.

\subsection{Training and performance}

\begin{figure}
  \begin{minipage}[r]{0.14\linewidth}
    \vspace{0.1cm}
    \begin{mybox}[black]\textbf{Ground truth}\end{mybox}
    \vspace{0.6cm}
    \begin{mybox}[black]\textbf{Prediction}\end{mybox}
    \label{fig:train_voidmask}
  \end{minipage}%
  \hfill
  \begin{minipage}{0.85\linewidth}
    \includegraphics[width=\textwidth]{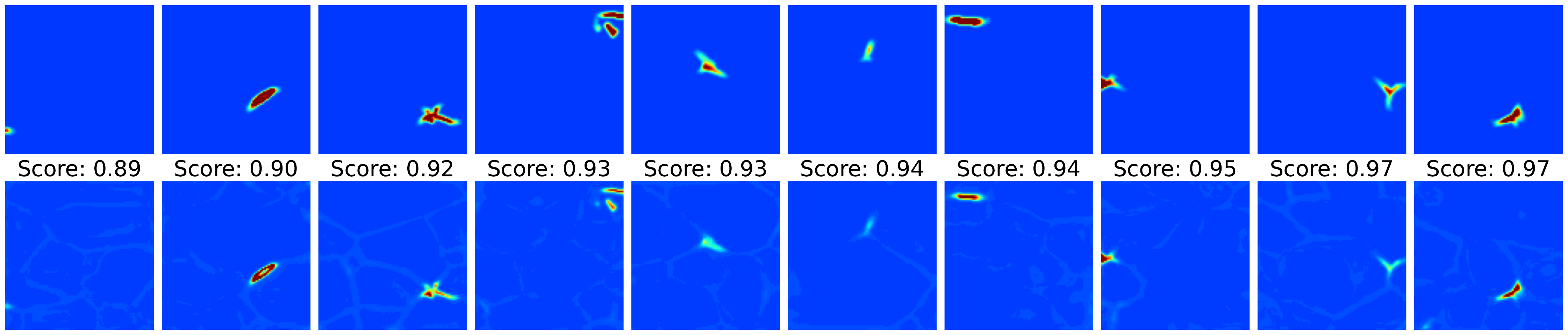}
  \end{minipage}
  \caption{Representative selection of model results for training sets, indicating the robustness of the ML architecture for capturing void nucleation probability fields. Intensity values are clamped between 0.05 and 0.6 for better visualization.}
  \label{fig:tr_voidmask}
\end{figure}

To train the model, the 57 {\it post-mortem} experimental microstructures are reconstructed and partitioned to arrays of 100x100, partially overlapping ``mini-microstructures'' that represent distinct neighborhoods around individual voids.
This cropping procedure produces a larger data set while preserving diverse microstructural contexts, thereby equipping the model with more comprehensive information about the regions surrounding voids.
All mini-microstructures that contain excessively large voids are discarded permanently.
To improve the balance of the dataset, all mini-microstructures that do not contain voids are discarded as well.

The model is tested and trained using ``N''-fold cross-validation.
Note that we use ``N'' (rather than ``k'') to emphasize the fact that each testing partition is comprised of all mini-microstructures from a single experimental micrograph, and as such, may have a variable number of associated mini-microstructures.
It is emphasized though this reduces model performance, it significantly enhances the model's reliability, as it shows good model performance across experimental measurements.
Unsurprisingly, the results are very accurate for training data (\cref{fig:tr_voidmask}).

Despite the use of attention gates and segmentation, a significant class imbalance still exists in the dataset.
To further remediate this issue, a specialized focal loss function is incorporated.
The focal loss modifies the standard binary cross-entropy to down-weigh well-classified examples and focus learning on hard negatives \cite{lin2017focal}.
The focal loss is defined as: 
\begin{equation}
\mathcal{L}_{\text{focal}}
= - \,\alpha \, (1 - p_t)^{\gamma} \, \log\!\bigl(p_t\bigr),
\qquad
\text{with}\;\;
p_t =
\begin{cases}
p      & \text{if } y = 1, \\[4pt]
1 - p  & \text{if } y = 0,
\end{cases}
\label{eq:focal-loss}
\end{equation}
where $p \in (0,1)$ denotes the predicted posterior probability for the positive class and $y \in \{0,1\}$ is the ground truth label. 
In this work, the choice of $\alpha = 0.5$ is chosen since it assigns a higher prior importance to the void pixels, while  $\gamma = 1$ is chosen since it sharply down-weighs well classified (easy) examples, directing the network’s attention toward hard, minority class pixels located at grain boundaries where small voids nucleate.

Using the focal loss as the loss function, a comparison is conducted between the standard U-Net and the Attention U-Net. 
For both the architectures, the training loss consistently decreases toward zero, indicating learning of the underlying mapping from microstructural features on the training set, with similar trends observed for all partitions. 
While both architectures exhibit similar training losses, the Attention U-Net achieves a lower testing loss, indicating improved generalization and better performance on unseen microstructural data (\cref{sec:accuracymetrics}). 
But the focal loss alone is insufficient to provide a complete assessment of the model performance.
Therefore a more conservative metric such as the Pearson Coefficient Correlation (PCC) is used to evaluate model performance. 
The average PCC values computed across the training and testing data demonstrate that the Attention U-Net achieves consistently higher average PCC values than the regular U-Net on both training and testing data indicating better model performance (\cref{sec:accuracymetrics}). 
A more in-depth analysis of the model performance using the PCC is discussed later (\cref{{sec:conservativemetric}}).
While the reduction in training loss indicates effective learning, it is essential to complement this measure with additional performance metrics to obtain a more comprehensive understanding of the model’s predictive behavior.

\subsection{Accuracy metrics}\label{sec:accuracymetrics}

\begin{figure}
\begin{subfigure}{0.5\linewidth}
  \centering
  \includegraphics[height=6.5cm]{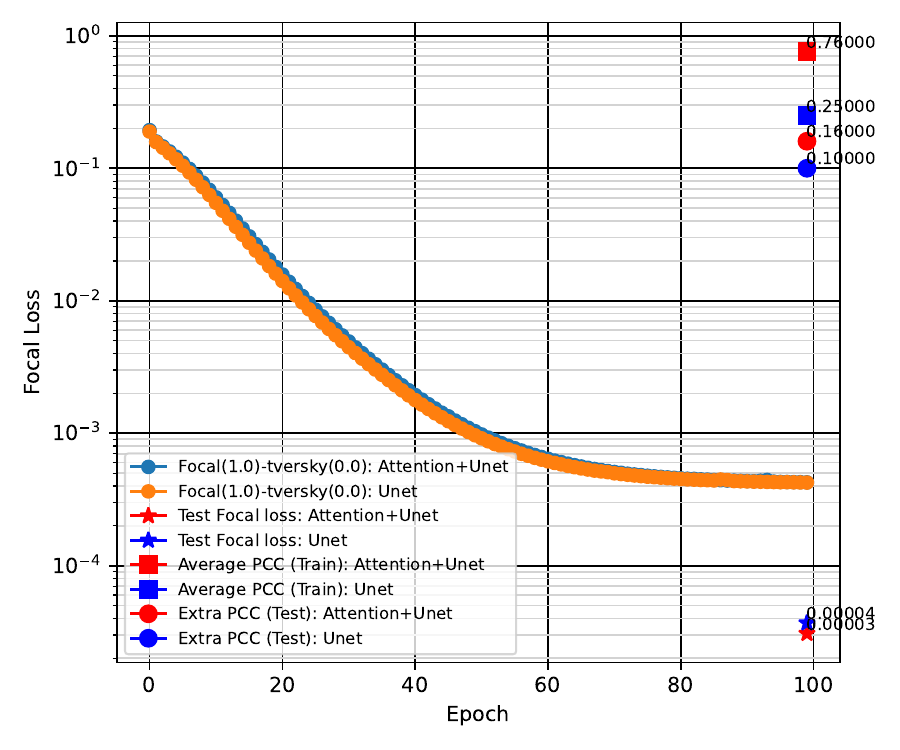}
  \caption{Focal loss vs number of epochs showing learning in a representative dataset.
  While both models achieve low testing loss values (with Attention U-Net performing better), focal loss alone does not provide a complete assessment of model performance. 
  }
  \label{fig:loss_curves}
\end{subfigure}\hfill
\begin{subfigure}{0.45\linewidth}
  \centering
  \includegraphics[height=6.5cm]{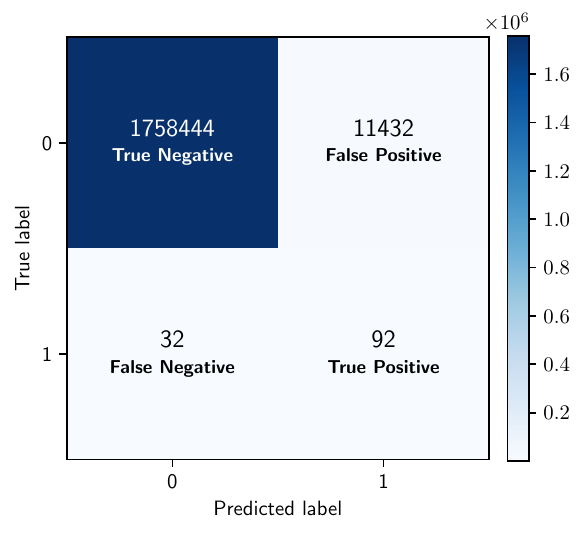}
  \caption{Pixel-wise confusion matrix.
  Most results are true negative, due to the sparsity of voids in the dataset.
  Many false positives result from the stochastic nature of nucleation.
  Importantly, very few false negatives are present.
  }
  \label{fig:confusion_matrix}
\end{subfigure}
\caption{Traditional model performance metrics}
\label{fig:traditional_metrics}
\end{figure}

Since focal loss alone does not provide a comprehensive evaluation of model performance, multiple statistics and metrics are considered to understand how the ML model performs during the testing process.
Many of the conventional metrics used in ML offer limited insights into the model's true behavior during the testing stage due to the class imbalance (for example, \cref{fig:traditional_metrics}), due to the imbalance and inherent stochasticity of nucleation behavior \cite{hahn2018role}.
To accommodate the imbalanced/stochastic nature of the problem, while maintaining accuracy, the following metrics are proposed.

\subsubsection{Conservative metric: Pearson correlation coefficient (PCC)} \label{sec:conservativemetric}

A broad range of traditional metrics used in U-Net was surveyed, including Hausdorff Distance \cite{cao2022swin}, pixel error \cite{ronneberger2015u} and the structural similarity index \cite{sharma2022deep}.
It was determined that most of these measures provide misleadingly high scores despite poor predictions made by the model.
To address this limitation, a more stringent and informative metric is required. 
The PCC \cite{benesty2009pearson}offers a robust alternative, as it quantifies the linear correlation between predicted and reference images, thereby providing a more reliable assessment of model performance.
The ground truth and prediction images are flattened into vectors X and Y respectively with the $PCC(X,Y)$ defined as:
\begin{equation}
PCC[X,Y]
= \frac{\mathrm{cov}(X,Y)}{\sigma_X\,\sigma_Y},
\end{equation}
where $\mathrm{cov}(X,Y)$ is the covariance, indicating the extent to which $X$ and $Y$ vary together relative to their means. 
The normalizing quantities $\sigma_X$ and $\sigma_Y$ are the standard deviations of $X$ and $Y$, respectively.
This normalization addresses the class imbalance, ensuring that $PCC$ is dimensionless and always bounded between $-1$ and $1$, where $+1$ represents a perfect positive linear correlation, $0$ represents no linear correlation, and $-1$ represents a perfect negative linear correlation.
It was determined that a value of $PCC>0.1$ indicates a reasonably good match.
(Additional details and comparison are provided in \cref{sec:discussion_pcc})

\subsubsection{Permissive metric: region-of-interest semimetric (ROI)}

Because of the stochasticity of the problem, it is expected that there are numerous false-positives (model indicates void likelihood where no void was observed) predicted by the ML model.
On the other hand, the presence of false-{\it negatives} (void present in a region not predicted by the model) is much more serious and is a better measure of the model's performance.
To explicitly highlight this aspect of model performance, the following metric is proposed:
\begin{align}
  \text{ROI}(X,Y) &= \frac{\langle X, Y \rangle_{\hat{\Omega}(X)}}{|X|_{\hat{\Omega}(X)}|Y|_{\hat{\Omega}(X)}}
  &
    \hat{\Omega}(X) &= \{\bm{x}\in \Omega : X(\bm{x})>\varepsilon\}
\end{align}
where $X,Y$ are the ground truth and predicted maps, respectively; $\Omega_{X}\subset\Omega$ is the ``region-of-interest'' in the ground truth {\it only} (the regions in which void nucleation activity is identified); $\varepsilon$ is a threshold for determining activation.
The operators $\langle\cdot,\cdot\rangle_{\hat{\Omega}(X)}, |\cdot|_{\hat{\Omega}(X)}$ are the standard L2 inner product and norm, respectively, over the ground truth region of interest.
It is noted that the restriction to a subset of $\Omega$, and the elimination of commutativity $ROI[X,Y]\ne ROI[Y,X]$ renders ROI a non-standard metric (hence ``semi-'' designation).

The ROI metric is effectively based solely on the (inverse) percentage of false negatives predicted by the ML model.
This is advantageous because low false negativity is a good indicator of proper model performance.
On the other hand, the ROI metric is obviously very permissive, as illustrated by the fact that substituting $Y=1$ maximizes the ROI metric for a given $X$.
Therefore, a combination of ROI and PCC is recommended for systematic evaluation of performance. 

\section{Results}\label{sec:results}

\begin{figure}
  %
  %
  \begin{subfigure}[b]{\textwidth}
    \begin{minipage}[r]{0.14\linewidth}
      \vspace{0.1cm}
      \begin{mybox}[black]\textbf{Ground truth}\end{mybox}
      \vspace{0.6cm}
      \begin{mybox}[black]\textbf{Prediction}\end{mybox}
    \end{minipage}%
    \hfill
    \begin{minipage}{0.85\linewidth}
      \includegraphics[width=\textwidth]{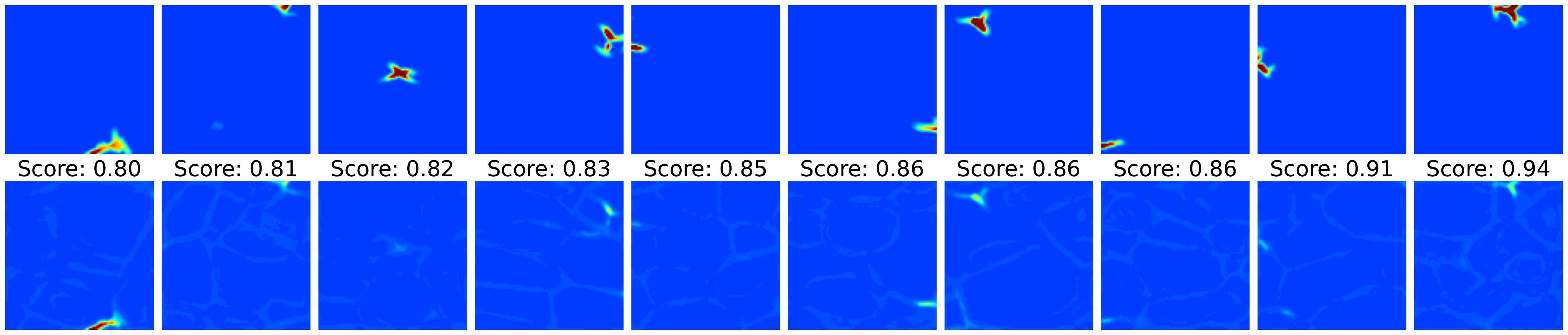}
    \end{minipage}
    \caption{High-scoring testing set (PCC>0.8))
    }
    \label{fig:test_highscore}
  \end{subfigure}
  \begin{subfigure}[b]{\textwidth}
    \begin{minipage}[r]{0.14\linewidth}
      \vspace{0.1cm}
      \begin{mybox}[black]\textbf{Ground truth}\end{mybox}
      \vspace{0.6cm}
      \begin{mybox}[black]\textbf{Prediction}\end{mybox}
      \label{fig:lowscore_voidmask}
    \end{minipage}%
    \hfill
    \begin{minipage}{0.85\linewidth}
      \includegraphics[width=\textwidth]{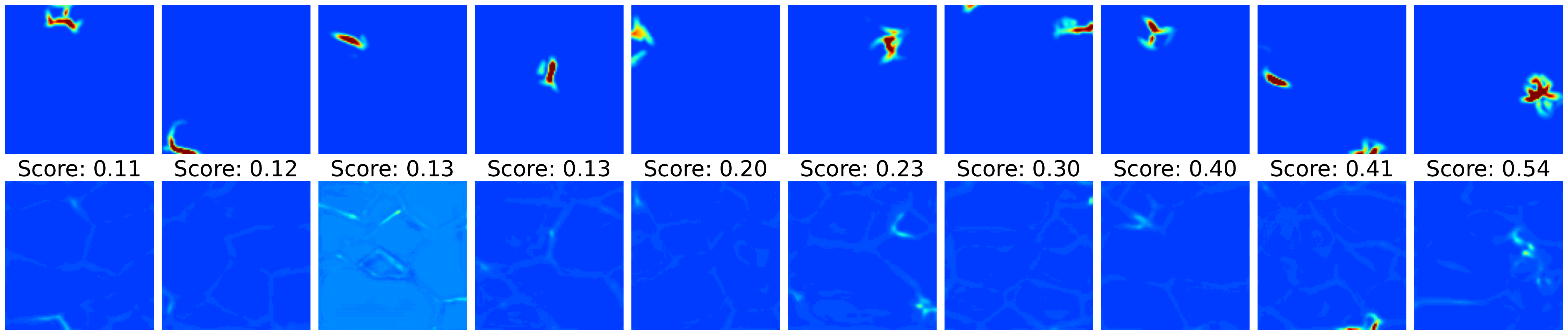}
    \end{minipage}
    \caption{Low-scoring testing results (0.8 > PCC > 0.1)
    }
    \label{fig:test_lowscore}
  \end{subfigure}
  \caption{
  Representative selection of model predictions for the testing sets (a-b).
  Color corresponds to no void nucleation likelihood (0 - blue) and high nucleation likelihood (1 - red). Intensity values are clamped between 0.05 and 0.6 for better visualization. 
  Results in (a/b) have a high/low degree of structural similarity to the ground truth, reflected in a high PCC value; however, the majority of actual nucleation sites are accurately captured. 
  }
  \label{fig:voidmask}
\end{figure}

Following the N-fold cross-validation strategy, the model is tested on each of the 57 micrographs where the model for each micrograph is trained exclusively on the remaining 56.
This ensures that the training results accurately reflect the model's performance on unseen experimental data, indicating confidence in its ability to make predictions based on future experiments.
The output is a predicted void nucleation map that can be compared directly to ground truth (\cref{fig:voidmask}).

The PCC metric is used to sort results into ``high-scoring'' (PCC>0.8, \cref{fig:test_highscore}) and ``low-scoring'' (0.8>PCC>0.1, \cref{fig:test_lowscore}).
Remarkably, for nearly all of the images, the ground-truth location of void nucleation sites is predicted nearly perfectly - even for low-scoring PCC images.
Visual inspection of the results shows that low PCC scores are primarily caused by (i) structural differences in the shape of the void nucleation regions, and (ii) the presence of ``medium-likelihood'' nucleation regions in the ML results.
Both of these discrepancies are clearly not problematic: it is naturally anticipated that the stochastic nature of the problem will statistically accurate nucleation maps, though not necessarily point-wise accurate.
Moreover, it is expected, and even desired, that the ML model yield probabilities for nucleation at other GBs - sites that potentially could have nucleated voids, yet didn't.
While this reasoning is admittedly speculative (by necessity - one is never told what {\it would have happened}), is an indication of accuracy that the model yields trends so consistently with expectation.

\begin{figure}
  \begin{subfigure}{0.49\linewidth}
    \centering
    \includegraphics[height=5cm]{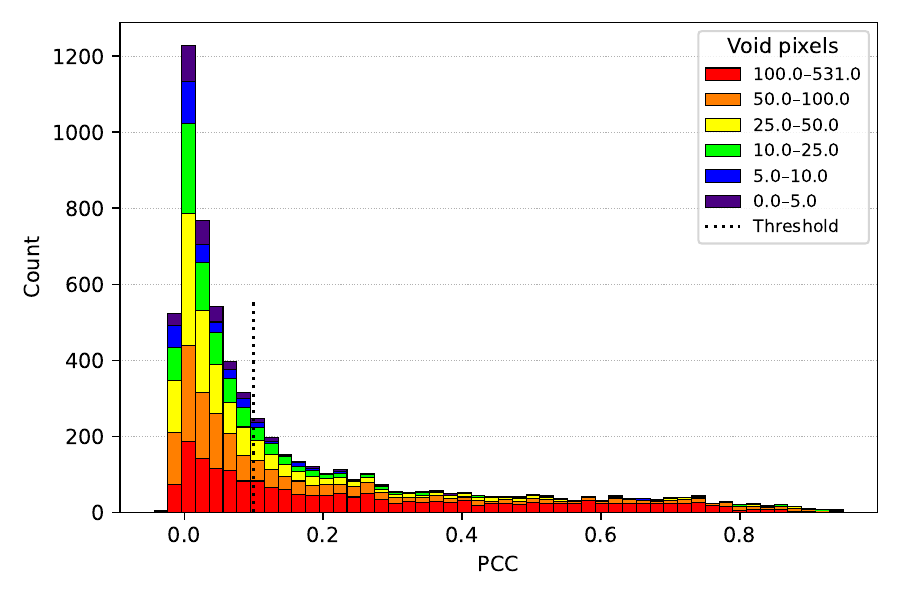}
    \caption{
    The PCC metric is used to compute the correlation between predicted and ground truth images across all 57 micrographs.
    Colors correspond to fraction of probability plots exceeding 0.3.
    The majority of low-scoring results are also fields with minimal nucleation activity in ground truth.
    }
    \label{fig:pearson_overall}
  \end{subfigure}\hfill%
  \begin{subfigure}{0.49\linewidth}
    \centering
    \includegraphics[height=5cm]{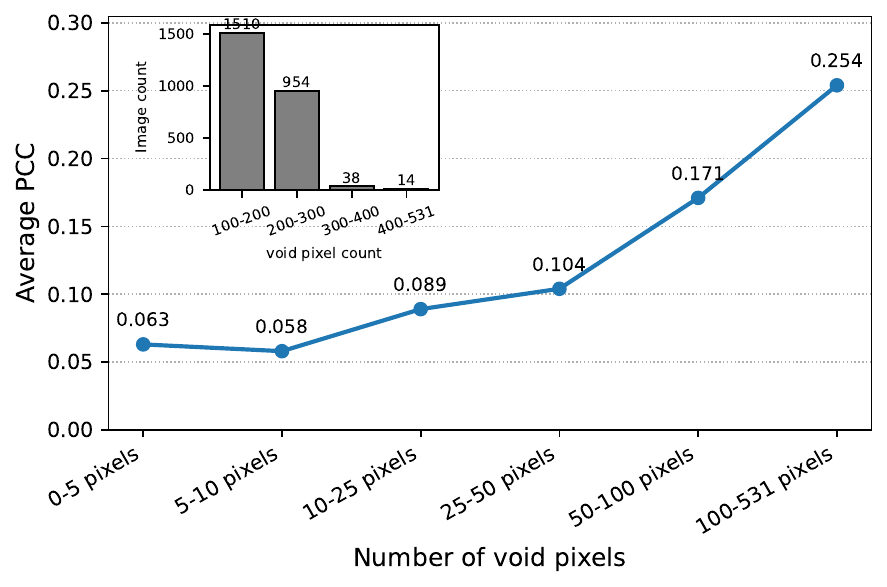}
    \caption{
    Correlation of PCC with ground truth nucleation activity.
    Maximum PCC is only 0.25 for the largest bin, representing meaningful performance.
    Other results fall into the 100–200 pixel sub-bin, illustrating the effect of imbalance and false positives on PCC.
    }
    \label{fig:avgpcc}
  \end{subfigure}
  \caption{PCC analysis of model performance across entire dataset}
\end{figure}

The training set exhibits consistently high PCC values, reflecting strong agreement between predictions and ground truth but the focus of this section lies on the testing set performance.
Across the test set, the PCC between ground truth and predictions predominantly range between 0 to 0.9 (\cref{fig:pearson_overall}).
For images with extremely sparse annotations of voids with about 0-5 void pixels out of a total of 10,000 pixels in the ground truth image (<=0.05 \% of the total image area), the PCC is centered around 0 with very few cases reaching even 0.5. 
A similar pattern holds for 5–10 void pixels (0.05–0.10\% of the total image area). 
When the ground truth contains 25–50 void pixels (0.25–0.50\% of the total image area) , the distribution remains centered near 0, but a larger fraction of cases surpass 0.5.
For 50–100 pixels (0.5–1.0\% of the total image area) and 100–531 pixels (1.0–5.31\% of the total image area), some images still yield near-zero correlations, yet many achieve substantially higher PCC values, even approaching ~0.9. 
Overall, these trends reflect the strong effect of class imbalance: as the prevalence of void pixels increases, the effective PCC correspondingly improves.
Importantly, low PCC values do not necessarily indicate poor predictive performance of the model and can be misleading under severe class imbalance.
In several cases with sparse void pixels, the model correctly localizes void regions but exhibits a low  PCC due to the presence of false positives and weak void predictions along different boundaries.

To understand the model behavior better, PCC values are computed for all images and then averaged within bins which are defined by the number of pixels associated with voids. 
The results indicate that average PCC increases with void size (which is indicated by the number of void pixels), demonstrating that the model performance during testing depends on the number of void pixels present (see \cref{fig:avgpcc}).
The maximum PCC reaches ~0.25 for the largest bin (100–531 pixels, corresponding to 1–5.3\% of the total image area), in spite of the severe class imbalance. 
Most images in this range fall within the 100–200 pixel sub-bin, corresponding to void areas of ~1–2\% of the image, highlighting that the severe imbalance persists and that the model seldom encounters large voids. 
The relatively low PCC values are further influenced by false positives and weak nucleation predictions along multiple grain boundaries which reflect the inherent stochasticity and physical complexity of void nucleation rather than purely model inconsistencies.

To guard against false impressions of performance, PCC is employed because, unlike SSIM, it does not spuriously reward trivial agreement under severe class imbalance. 
When an image with a minute void fraction is compared with an image containing no voids, SSIM can remain high due to the similarity of background pixels, whereas PCC remains near zero, reflecting the lack of correlation in the sparse void signal. 
This behavior is verified empirically: for representative ground-truth images, PCC is computed against the model’s predictions, against randomized predictions, and against a randomly generated fields.
Low PCC values are obtained for the random baselines, and elevated values emerge only when genuine spatial correlation is present.
Accordingly, PCC provides a more stringent and discriminative assessment of prediction quality and avoids the false highs that can arise with SSIM and other popular metrics under extreme class imbalance.

\begin{figure}
  \begin{subfigure}{0.25\linewidth}
    \centering
    \includegraphics[height=6cm]{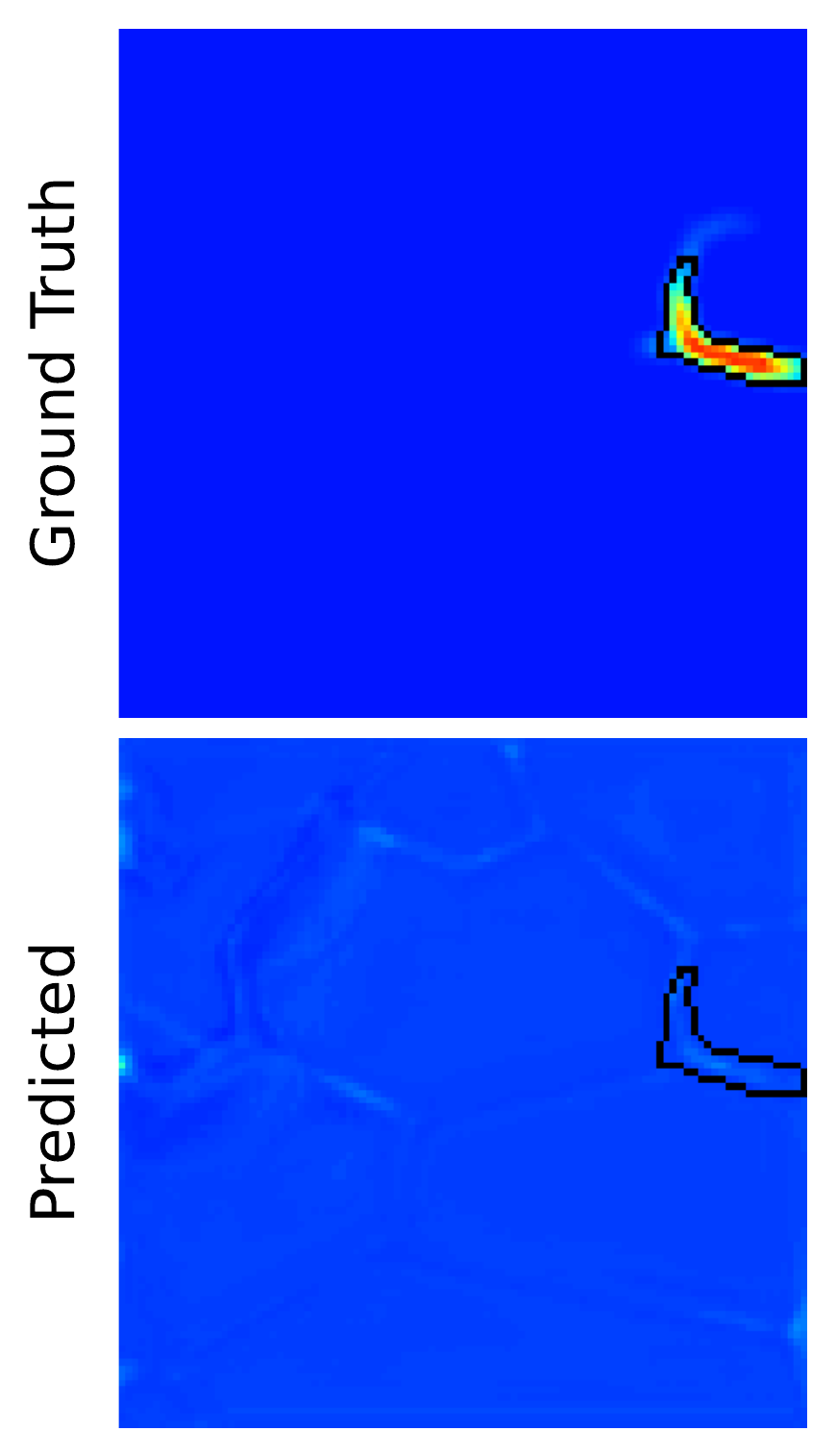}  
    \caption{Graphical illustration of ROI metric: the region of interest around a void is projected onto the predicted image.}
    \label{fig:roi}
  \end{subfigure}\hfill
  \begin{subfigure}{0.73\linewidth}
    \centering
    \includegraphics[height=6cm]{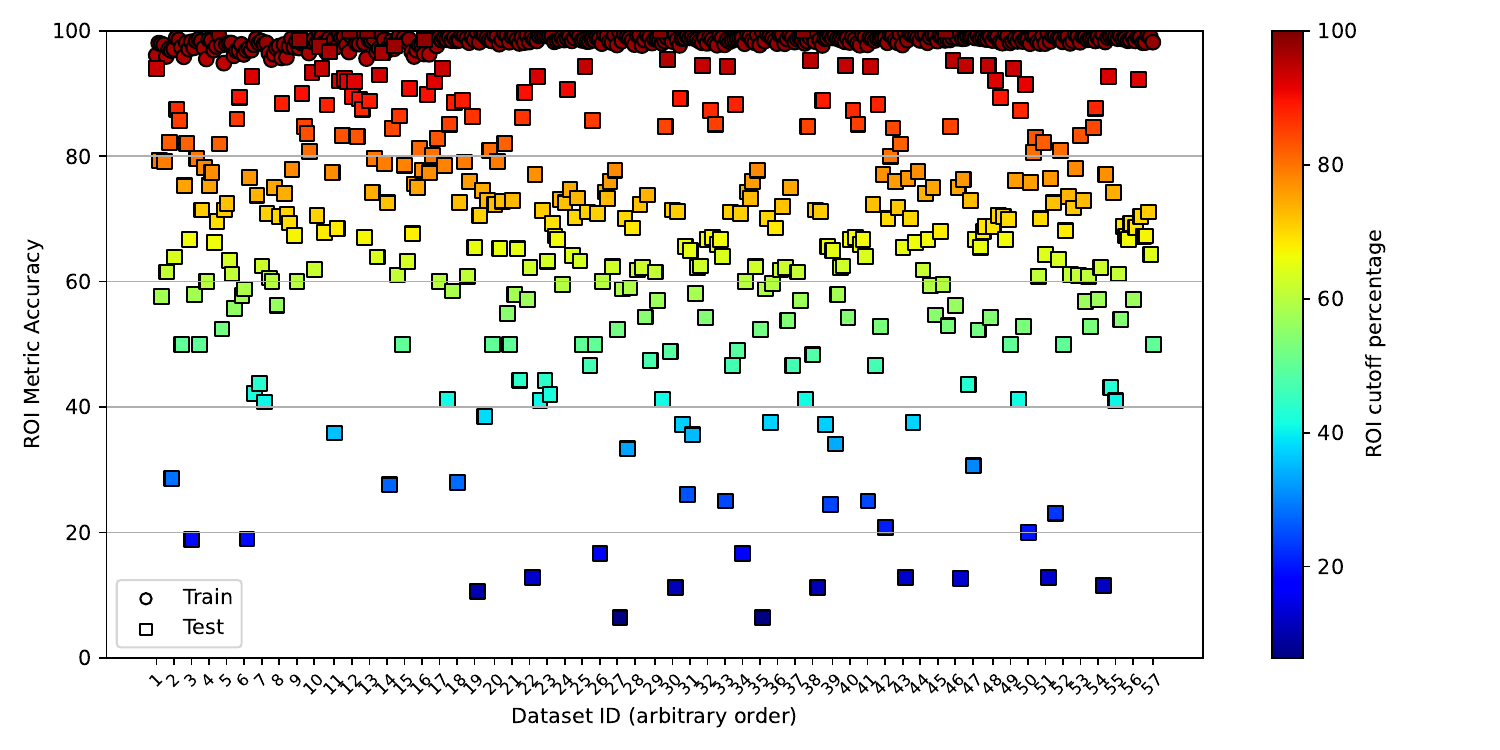}
    \caption{
    Cross-validation across all 57 micrographs
    Each train/test pair corresponds to the model results for a specific image, tested against a model trained using all other images as training data (x-axis corresponds to the image index, which is arbitrary).
    Color corresponds to the ROI threshold.
    }
    \label{fig:metric}
  \end{subfigure}
  \caption{ROI metric and cross-validation of model}
  \label{fig:pcc_eval}
\end{figure}

ROI analysis is performed for all datasets from the N-fold cross validation.
The analysis is performed using various thresholds to understand model behavior.
The training accuracies remain consistently above 90\% across all thresholds,, confirming that the network accurately identifies void locations during training. 
The testing accuracies exhibit variability in accuracy depending on the threshold, with the majority of values concentrated in the 70\% range (\cref{fig:pcc_eval}).
Even though the training accuracy is over 90\% for all datasets, it should be noted that it is not trivial for the model to predict voids from grain orientation and grain boundary energy information, although it can see target masks in the loss calculation. 
This indicates that the model effectively predicts void nucleation sites.
The model does not see ground truth images during the testing stage, but the model does surprisingly well on a majority of datasets.
This indicates that the model is actually able to learn and predict where voids could nucleate from grain orientations and grain boundary energy information which is very encouraging.

\section{Conclusion}\label{conclusions}
In this work, a U-Net architecture augmented with attention gates has been developed and evaluated for predicting void nucleation in microstructural images, using multiple input channels representing grain boundary energy and grain orientation. 
In addition to explicitly incorporating these channels, the use of orientation images implicitly provides information about grain size, misorientation, and grain aspect ratio, thereby enriching the feature space available to the model. 
By employing focal loss, the model demonstrates robustness under severe class imbalance, where the majority of pixels correspond to non-void regions. 
The training loss converges to low values, confirming effective learning during training.  
Analysis of the pixel-wise confusion matrix shows that the model achieves a high true negative rate, indicating that regions with no voids are well captured.
Although a higher false positive rate is observed, these predictions may reflect potential void nucleation sites, consistent with the inherently stochastic nature of void nucleation. 
Further, a good true positive rate shows that the model captures void regions well.
There is also some false negative rate which indicates that the model fails to capture some void regions.

Qualitative comparisons across training and testing samples further demonstrate that the model reliably identifies void nucleation sites along grain boundaries, while also highlighting additional regions that could be susceptible to future void formation. 
Although labeled as false positives, these predictions may serve as early indicators of potential void nucleation sites, capturing the stochastic behavior of void nucleation under spallation.  
Quantitative evaluation with the PCC highlights the challenges imposed by extreme class imbalance.
While PCC values on the training set are consistently high, test set values span the range $0$–$0.9$, with performance strongly dependent on void size. For images with extremely sparse annotations ($<0.1\%$ of the total image area, corresponding to 0–10 void pixels), average PCC values across images are centered near zero, reflecting the difficulty of reliable correlation under such imbalance. 
As the number of void pixels increases, the average PCC improves, reaching a maximum of $\sim$0.25 in the largest bin (100–531 void pixels, or $1$–$5.3\%$ of the total image area). 
Although modest in magnitude, these values are meaningful given the rarity of void pixels in the dataset.
Importantly, low PCC values in sparse cases do not necessarily indicate poor predictive ability: qualitative analysis reveals that the model often localizes voids correctly but appears penalized by false positives and weak nucleation predictions along additional grain boundaries. 
Such false positives may, in fact, capture the inherent stochasticity of void nucleation and highlight regions susceptible to future void nucleation.  

Despite these promising results, some limitations remain. 
The most significant is the reduced performance on the testing set, largely due to the limited size of the available experimental data.
Expanding the dataset with more varied microstructural examples is expected to improve generalization and predictive accuracy.
Another limitation arises from the ground truth probability field generation process: pixel-wise multiplication with the grain boundary energy map causes voids within grain interiors to vanish, restricting the model to nucleation predictions only along grain boundaries. 
Furthermore, the current framework assumes equal nucleation likelihood across a grain boundary, whereas regions such as triple junctions may present higher susceptibility. 
Incorporating spatial weighting along grain boundaries or region-specific priors could enhance the physical accuracy of predictions.
Also, it should be noted that the micrographs in this work were captured in a single shot. 
Acquiring multiple datasets under varied loading conditions and microstructural modifications would be extremely beneficial. 

Importantly, this work demonstrates {\it that criteria for void nucleation can be inferred from microstructure geometry and properties}.
While it is still not possible to detect a simple or reduced-order relationship here, this validates the long-held perspective that void nucleation behavior is (up to stochasticity) predictive.

\section{Acknowledgements}

BR, ATA and JP acknowledge support from the National Science Foundation, grant number 2341922.
SF acknowledges the support from the US Department of Energy through the Los Alamos National Laboratory.
Los Alamos National Laboratory is operated by Triad National Security,  LLC, for the National Nuclear Security Administration of U.S. Department of Energy (Contract No. 89233218CNA000001)
Finally, the authors wish to acknowledge Nihal Pushkar for fruitful discussions, and especially for suggesting the use of an attention gate mechanism.

\section*{Data Availbility}

All data and code used for this paper are available upon request. 

\bibliography{main}

\newpage

\appendix

\section{Dataset preparation}\label{sec:dataset_preparation}

\begin{figure}[h]
    \centering
    \begin{subfigure}[t]{0.3\textwidth}
        \includegraphics[width=\textwidth]{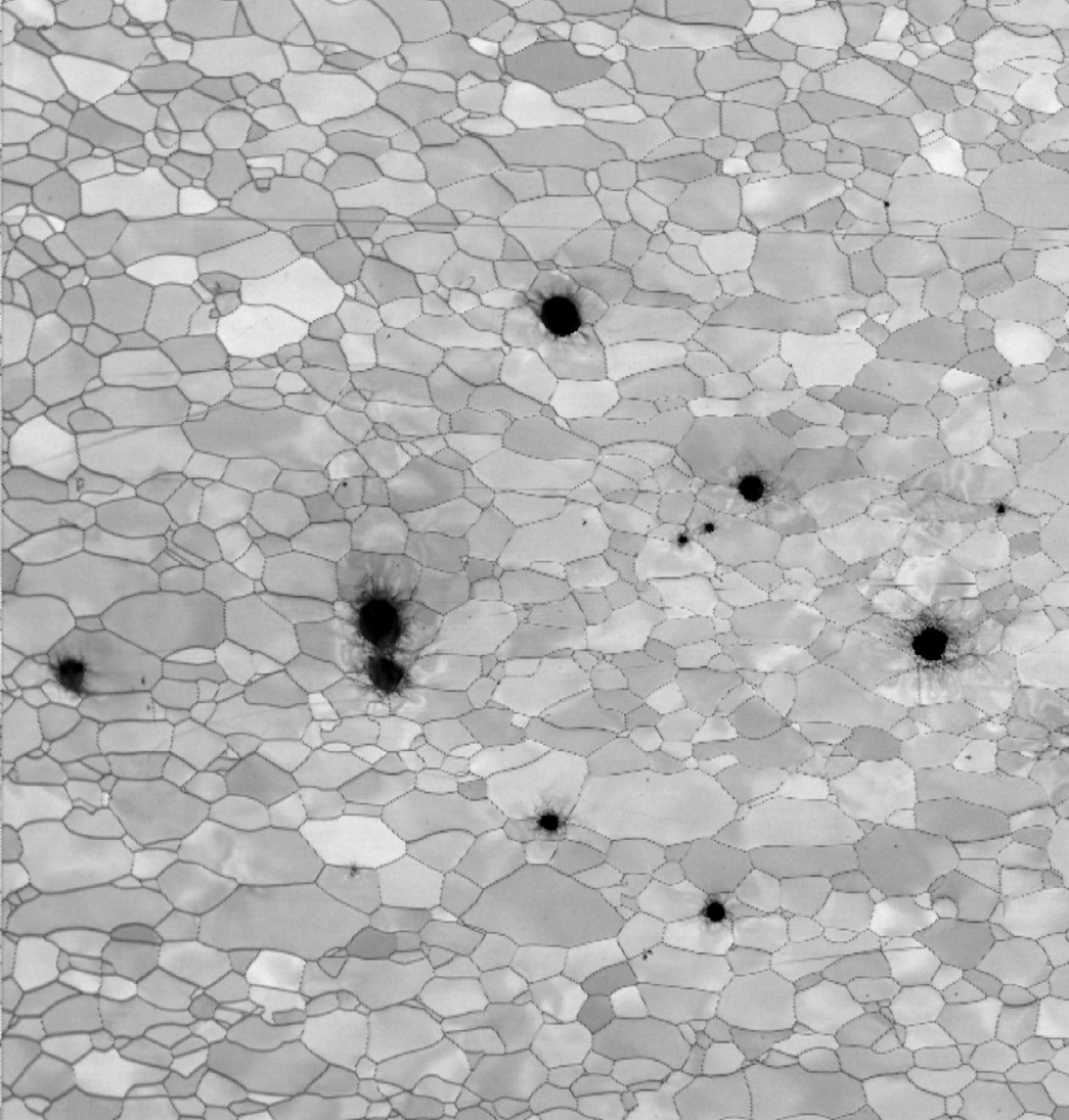}
        \caption{Original micrograph image of the post-incipient spalled microstructure of BCC tantalum}
        \label{subfig:voids}
    \end{subfigure}\hfill
    \begin{subfigure}[t]{0.3\textwidth}
        \includegraphics[width=\textwidth]{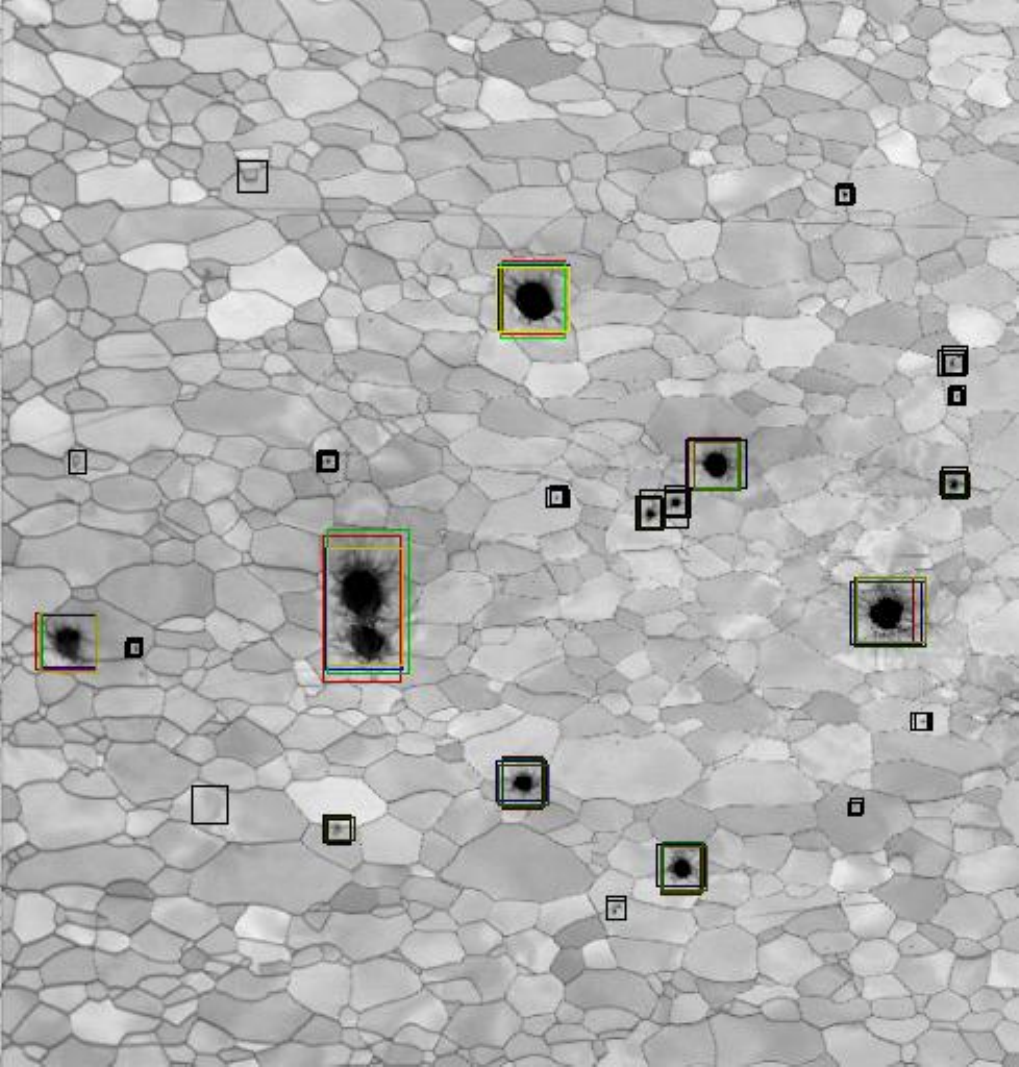}
        \caption{Identification of voids; more brightly colored boxes indicate void regions with higher confidence values}
        \label{subfig:void_identification}
    \end{subfigure}\hfill
    \begin{subfigure}[t]{0.3\textwidth}
        \includegraphics[width=\textwidth]{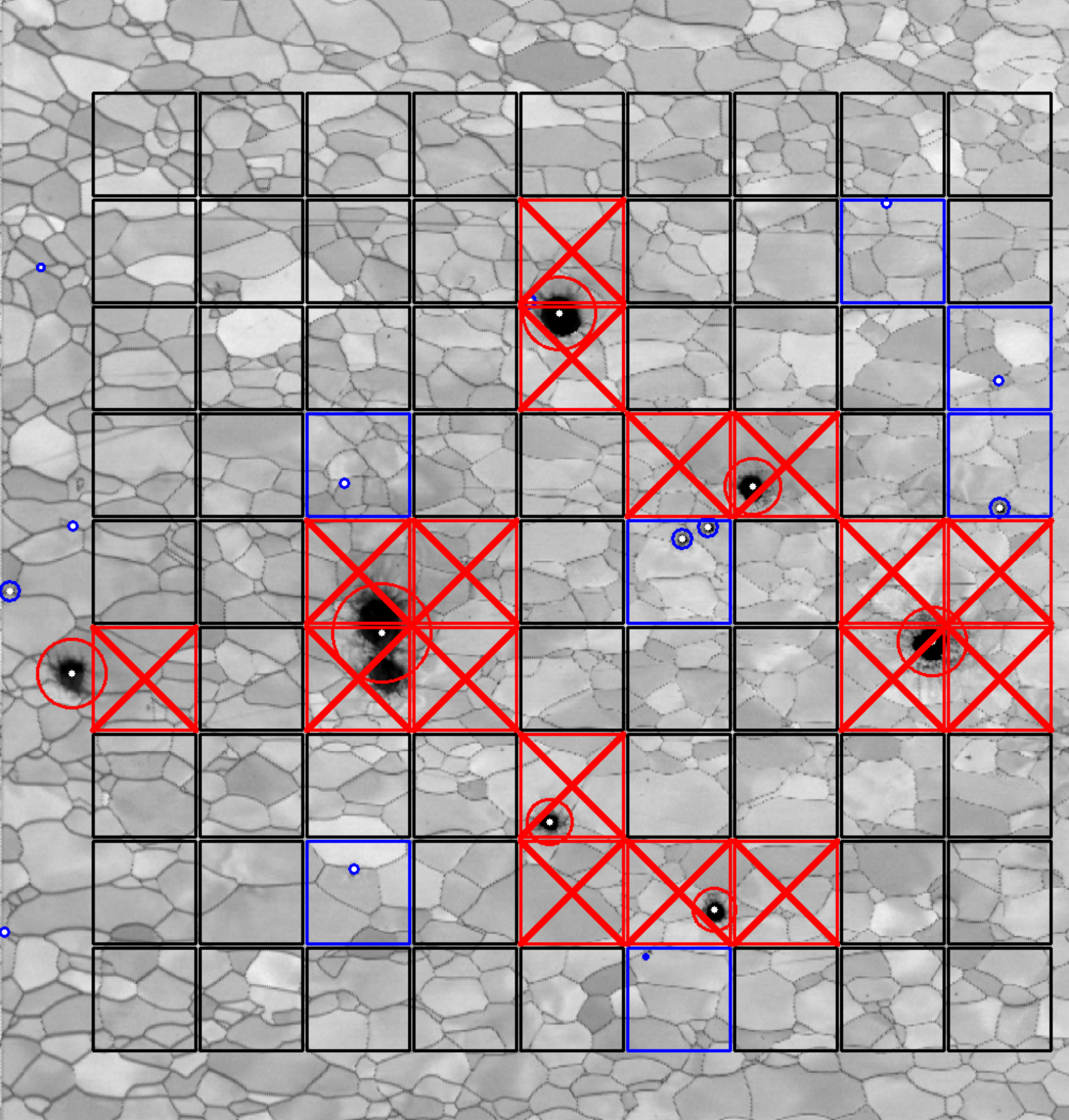}
        \caption{Micrograph segmented into red (bad voids), blue (good voids), and black (no void regions).}
        \label{subfig:segmentation}
    \end{subfigure}
    
    \vspace{1.5em}  

    \begin{subfigure}[t]{0.3\textwidth}
        \includegraphics[width=\textwidth]{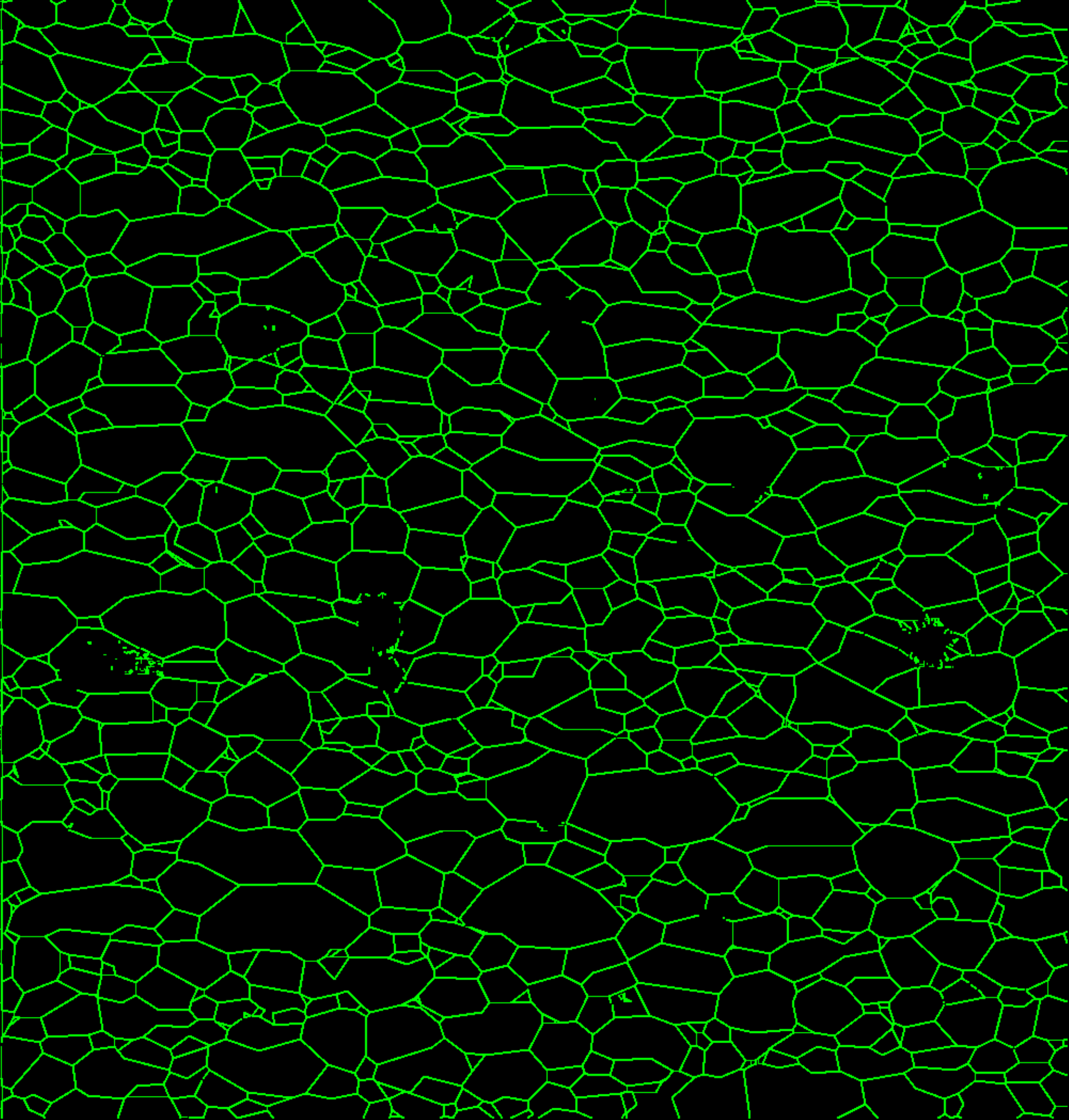}
        \caption{Grain boundary generation with the help of EBSD data}
        \label{subfig:EBSD}
    \end{subfigure}\hfill
    \begin{subfigure}[t]{0.3\textwidth}
        \begin{overpic}[width=\textwidth]{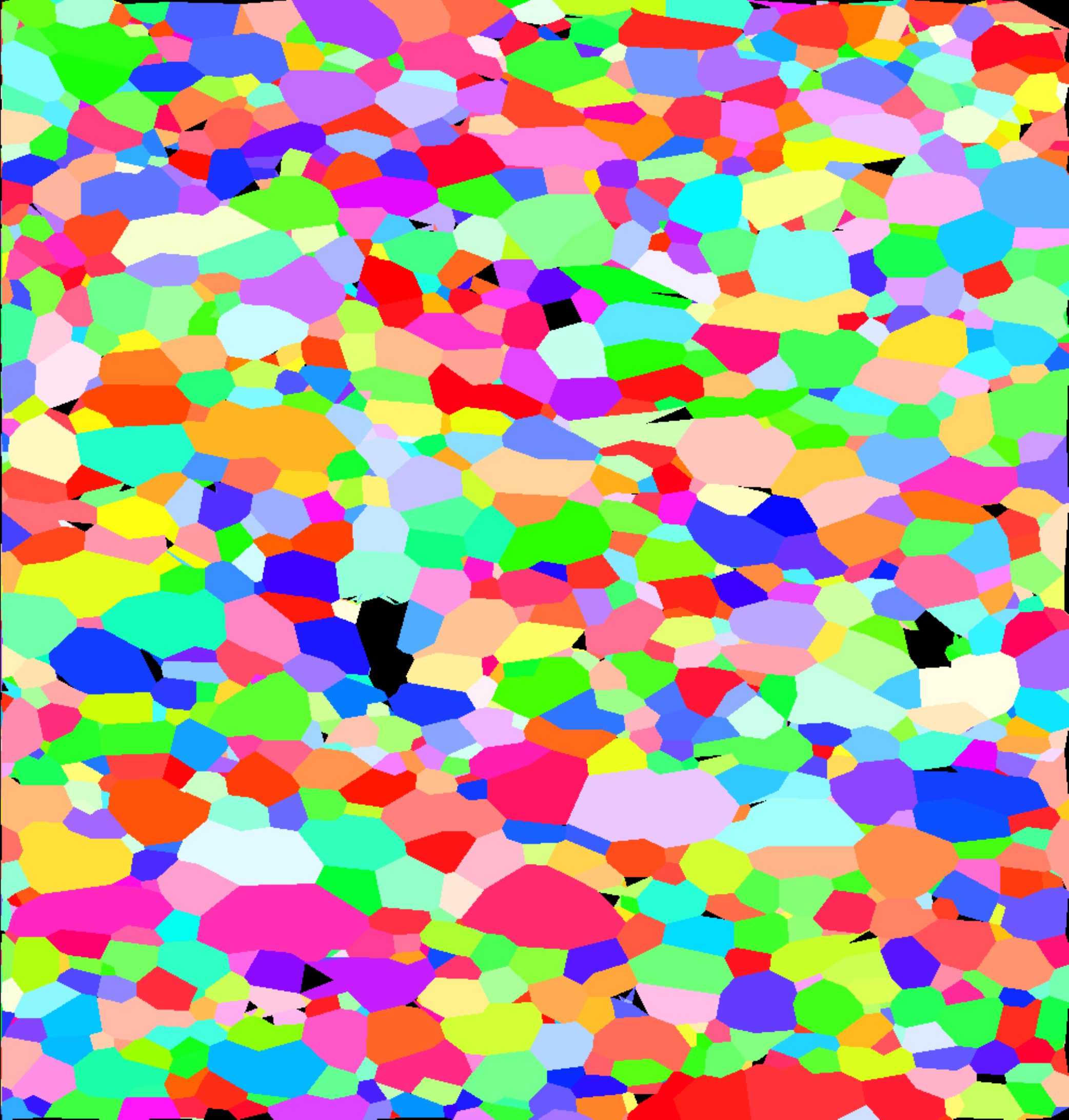}
          \put(00,0){\includegraphics[width=1.5cm]{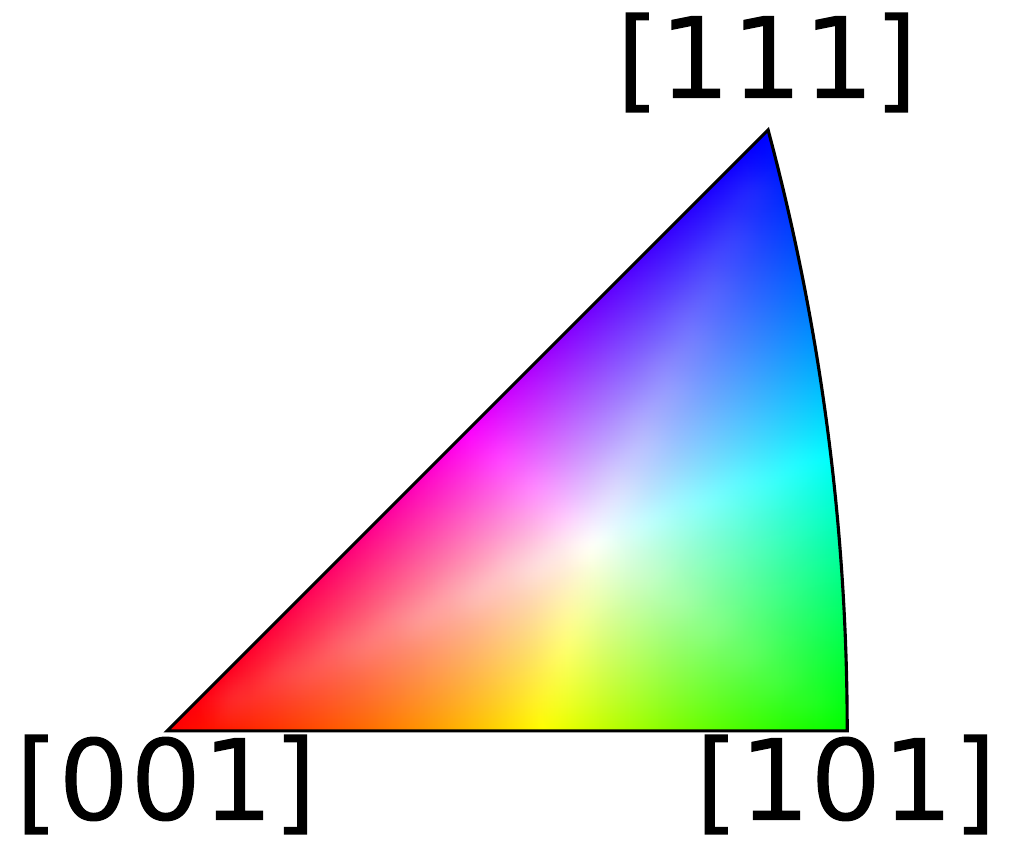}}
        \end{overpic}
        \caption{Grain reconstruction using EBSD data}
        \label{subfig:draw_grains}
    \end{subfigure}\hfill
    \begin{subfigure}[t]{0.3\textwidth}
        \includegraphics[width=\textwidth]{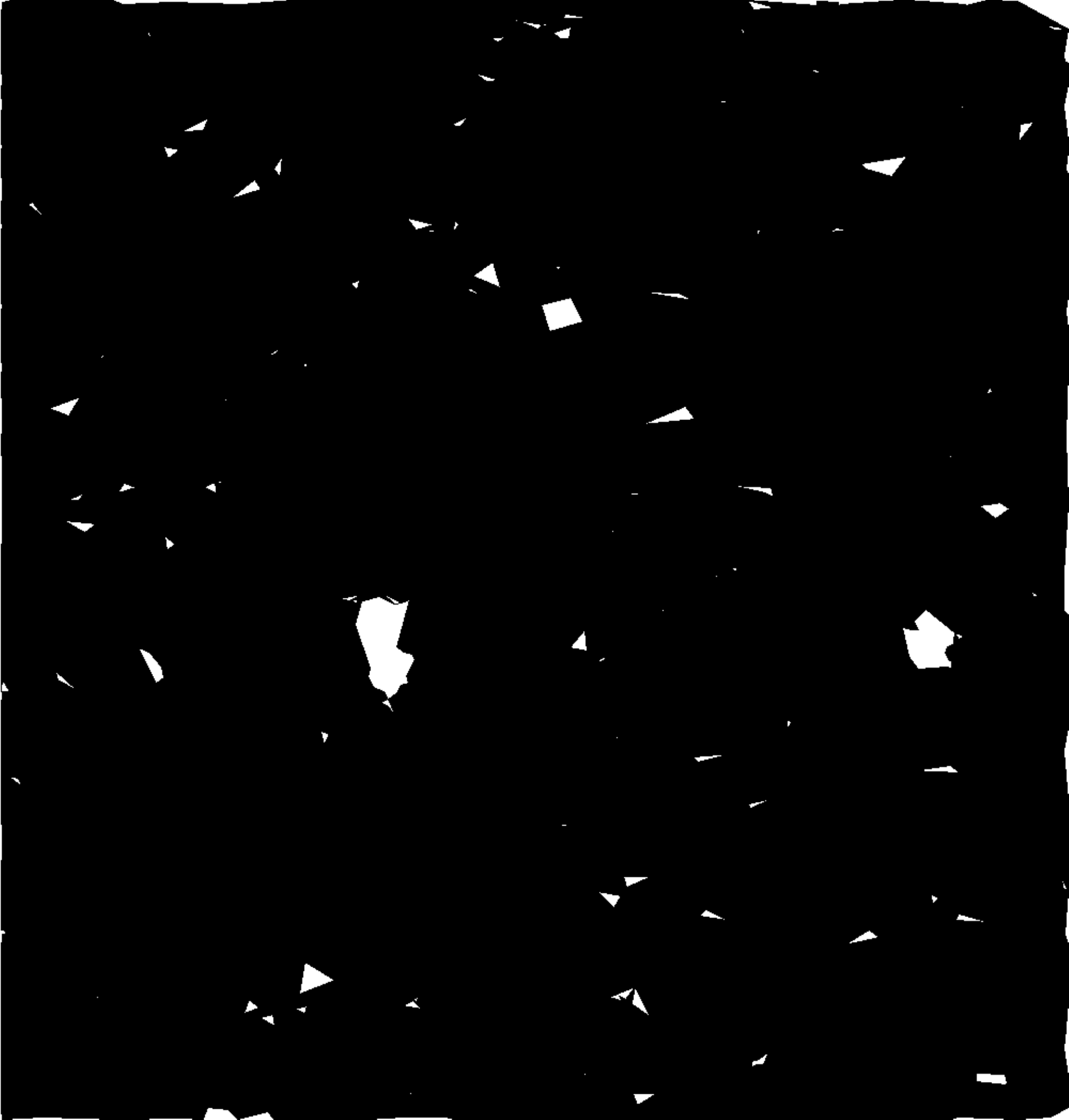}
        \caption{Mask applied to the microstructure to highlight regions of voids}
        \label{subfig:mask}
    \end{subfigure}

    \vspace{1.5em} 

    \hspace{5pt}
    \hspace{5pt}

    \caption{Combined overview of void identification, segmentation, and boundary reconstruction steps for microstructures. 
    Subfigures (a)--(c) show the original micrograph (a), identified voids (b), and segmented micrograph (c). 
    Subfigures (d)--(f) illustrate boundary generation using EBSD (d), grain reconstruction (e), and application of a mask (f). 
    }
    \label{fig:combined_voids_grains}
\end{figure}

The initial step of MRC focuses on identifying voids in {\it post-mortem} micrographs. 
This void detection step relies on an ML model that is trained using an object detection ML model called YOLOv5 \cite{ultralytics2021yolov5} to locate voids within microstructural images of BCC tantalum (see \cref{subfig:void_identification}).
The model is trained on manually classified micrographs and demonstrates high accuracy in both classifying voids and placing bounding boxes around them.
To improve detection accuracy, each micrograph is divided into 100x100 pixel segments, which are analyzed in four rotations (0°, 90°, 180°, 270°), and the resulting bounding boxes are consolidated. 
Each tile receives a score based on the confidence of the predicted voids across these rotations, with higher scores indicating the presence of voids. 
A score-based thresholding strategy is employed to classify segments as “void” or “no void,” with ambiguous cases left unclassified.
Following the identification of voids, they are classified into "acceptable" and "unacceptable" categories based on their radii (see \cref{subfig:segmentation}). 
Voids that exceed a critical size threshold are associated with significant plastic distortion, and cannot be completely reconstructed using MRC.
Such segments are excluded from the dataset. 
Based on this criterion, micrographs are segmented into red regions (unacceptable voids), blue regions (acceptable voids), and black regions (no voids) (see \cref{subfig:segmentation}). 
Only blue and black regions are retained for subsequent analysis, while red regions are discarded due to excessive microstructural damage.

\section{Discussion on PCC effectiveness}\label{sec:discussion_pcc}

\begin{figure}[h!]
  \centering
  \includegraphics[width=0.9\linewidth]{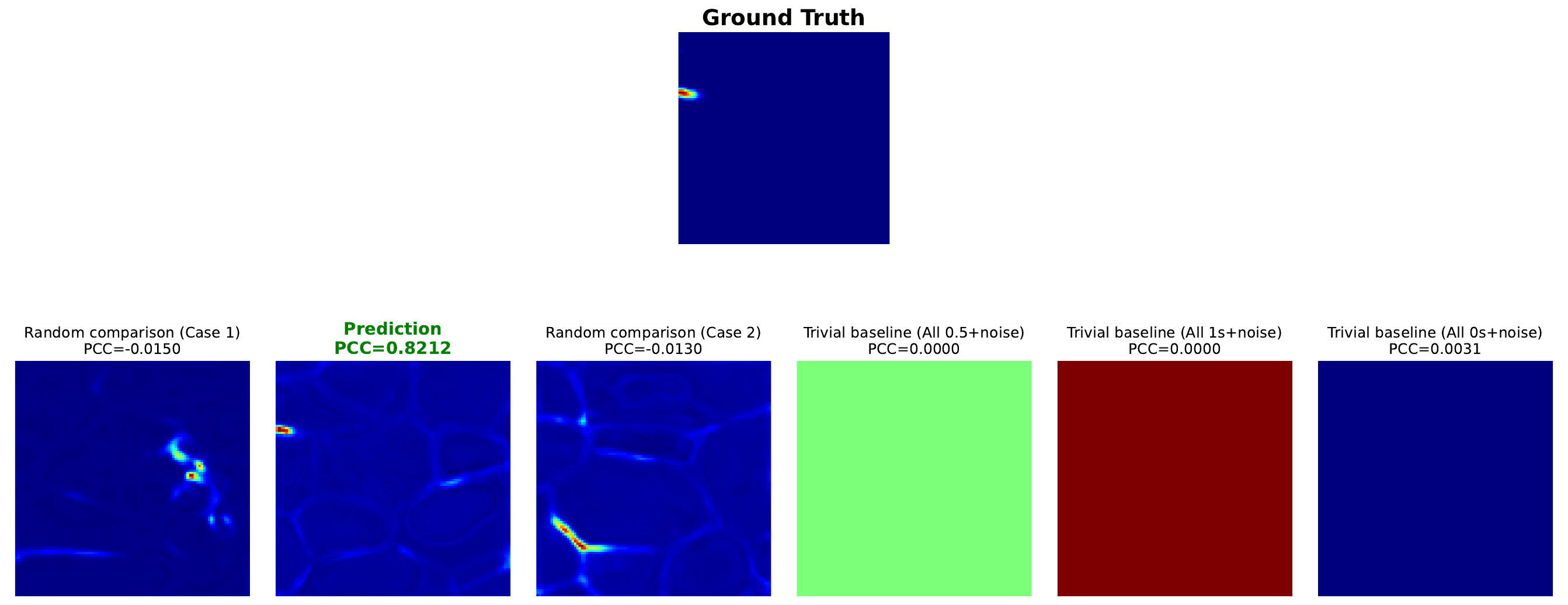}
  \caption{Pearson correlation coefficient based comparison between the ground truth and predicted void nucleation fields. The reference ground truth image is shown in the top row indicating the baseline against which all comparisons are made. The bottom row presents various predicted images, each annotated with its corresponding PCC value. The predicted field corresponding to the ground truth reference yields the highest correlation values as showm in green, while random predictors (such as constant fields of ones, zeros, or 0.5 values) exhibit significantly lower PCC values, underscoring their lack of structural correspondence. This shows that PCC is stringent as compared many other metrics which can give a false impression that the model predicts well even in cases where it does not do well.
  }
  \label{fig:metric_metric}
\end{figure}

This section evaluates PCC effectiveness by considering a representative ground truth evaluated against a variety of predcitions (\cref{fig:metric_metric}).

\end{document}